\documentclass[a4paper,10pt]{{scrartcl}}

\usepackage[utf8]{inputenc}
\usepackage[bookmarksopen,colorlinks,linkcolor=black,citecolor=blue]{hyperref} 

\usepackage{graphics}
\usepackage{graphicx} 
\usepackage{subfigure}
\usepackage{a4wide}
\usepackage{tabularx}
\usepackage{longtable}
\usepackage{wallpaper}
\usepackage{footnote}
\usepackage{multirow}
\DeclareGraphicsExtensions{.pdf}

\usepackage[citestyle=numeric 
	    , backref=false
	    , abbreviate=true
	    , uniquename=init
	    , firstinits=true
	    , terseinits=false
	    , bibstyle=numeric
	    , maxcitenames = 2
	    , maxbibnames = 99
	    , minbibnames = 3
	    , doi=false
	    , isbn=false 
	    , sorting = none 
	    , url=false
	    , natbib=true 
	    , backend=bibtex8
	    , useprefix=true]{biblatex}	    
\DeclareNameAlias{sortname}{last-first}
\DeclareFieldFormat[article]{volume}{\bibstring{}~#1}
\DeclareFieldFormat[article]{pages}{\bibstring{}~#1}

\hypersetup{linkcolor=blue,urlcolor=blue}

\renewbibmacro{in:}{
  \ifentrytype{article}{}{
    \printtext{\bibstring{in}\intitlepunct}
  }
}

\renewbibmacro*{volume+number+eid}{%
  \printfield{volume}%
  \ifpunctmark{,}{\addcomma}{\addcomma}
  \setunit*{\addcolon}
  \printfield{number}%
  \setunit{\addcomma}%
  \printfield{eid}
}

\AtEveryBibitem{%
  \clearfield{number}
  \clearfield{day}%
  \clearfield{month}%
  \clearfield{endday}%
  \clearfield{endmonth}%
}

\usepackage[numbered,framed]{matlab-prettifier} 
\usepackage{soul} 

\usepackage{pdfpages}
\usepackage[withpage]{acronym}
\usepackage{enumitem}
\usepackage{amsmath}
\usepackage{amssymb}
\usepackage[bottom]{footmisc} 
    
\title{\Large{Calibration and reference simulations for the \\auditory periphery model of Verhulst et al. 2018 version 1.2}}
\author{\large{Alejandro Osses Vecchi and Sarah Verhulst\footnote{\hspace{6pt}Contact: Alejandro Osses Vecchi (\href{mailto:alejandro.osses@ugent.be}{alejandro.osses@ugent.be}), prof. Sarah Verhulst (\href{mailto:s.verhulst@ugent.be}{s.verhulst@ugent.be}). \vspace{-20pt}}}\\ 
		\normalsize{Hearing Technology Lab @ WAVES,}\\
		\normalsize{Department of Information Technology, Ghent University}} 
\date{}

\begin{document}
\maketitle

\vspace{-40pt}
\begin{table}[!h]
	\centering
\scalebox{0.8}{
	\begin{tabular}{|llll|} \hline
	\multicolumn{4}{|l|}{\textbf{Abbreviations}} \\ 
	ABR & Auditory Brainstem Response & IC  & Inferior Colliculus          \\
	AM & Amplitude Modulation         & IIR & Infinite Impulse Response    \\ 
	AN & Auditory Nerve               & MTF & Modulation Transfer Function \\
	CF & Characteristic Frequency     & NH  & Normal hearing               \\
	CN & Cochlear Nucleus             & HI  & Hearing impaired             \\
	EFR & Envelope Following Response & pe SPL& peak-equivalent Sound Pressure Level\\  
    FFT & Fast Fourier Transform      & W-I,-III,-V & Waves I, III, V \\ 	\hline
	\end{tabular}}
\end{table}

\small{The code of the auditory periphery model can be retrieved from: \href{https://github.com/HearingTechnology/Verhulstetal2018Model/}{https://github.com/HearingTechnology/ Verhulstetal2018Model/}. The current model version can be cited as follows: ``We adopted the model of the human auditory periphery described by Verhulst, Altoè, and Vasilkov (\citeyear{Verhulst2018a}) in the v1.2 implementation (Osses~Vecchi \& Verhulst, 2019).''}

\vspace{-6pt}
\section*{Abstract}

This document describes a comprehensive procedure of how the biophysical model published by Verhulst et al. (2018) \cite[][]{Verhulst2018a} can be calibrated on the basis of reference auditory brainstem responses (ABRs). Additionally, the filter design used in two of the model stages, cochlear nucleus (CN) and inferior colliculus (IC), is described in detail. These descriptions are valid for a new release of the Verhulst et al. model, version 1.2, as well as for previous versions of the model (version 1.1 or earlier), whose block diagram is shown in Fig.~\ref{fig:diagram}  \cite[see also,][their Fig.~1]{Verhulst2018a,Osses2019b}. The differences between the model versions are explicitly mentioned  and simulations to basic auditory stimuli are shown for model versions 1.1 and 1.2. Version~1.2 of the model includes a new implementation of the CN and IC stages (Stages 5 and 6). All previous model stages (Stages 1-4 in Fig.~\ref{fig:diagram}: outer and middle ear, transmission-line cochlear filter bank, inner hair cell model, and auditory nerve model) remained unchanged.\vspace{4pt}

In the new release (model version 1.2), in addition to the updated CN and IC stages, we employed a different calibration procedure to match human reference ABR amplitudes of waves I, III, and V more faithfully. This release note shows the implications of these model adjustments on the simulations presented in the original 2018 model paper. For this purpose, results from two model versions are reported:

\begin{itemize}[leftmargin=*]
	\item New model release (version 1.2), labelled as \textbf{`model v1.2'} (GitHub commit \href{https://github.com/HearingTechnology/Verhulstetal2018Model/releases}{f59309b}). 
	\item Previous model release (version 1.1) as used in \cite{Verhulst2018a}, labelled as \textbf{`model v1.1'} (GitHub commit \href{https://github.com/HearingTechnology/Verhulstetal2018Model/releases}{9a7a7ef}).
\end{itemize}

The main difference between IC model stages relates to the degree of IC inhibition that was applied, with more inhibition in \textbf{v1.2} than implemented in \textbf{v1.1}. The time domain simulations presented in this document show that this change in inhibition strength does not drastically change the results presented in the original paper. However, \textbf{v1.2} more correctly captures the physiologically derived CN and IC inhibition/excitation strengths \cite{Nelson2004}. 

\section{Model description}

\begin{figure}[h!]
	\centering 
		\includegraphics[width=.95\textwidth]{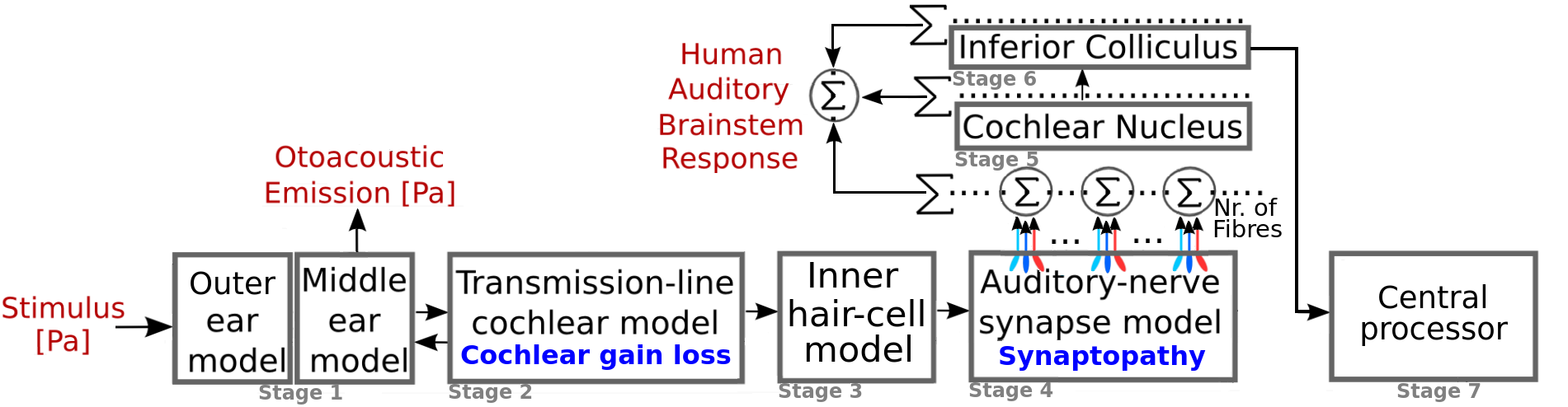}
		\vspace{-12pt}
		\caption{Block diagram of the biophysical model of the auditory periphery by Verhulst et al.~\cite{Verhulst2018a,Osses2019b}.}
		\label{fig:diagram}
\end{figure}

The model of the auditory periphery, whose framework is described in \cite{Verhulst2018a,Verhulst2015, Verhulst2012}, offers a quantitative description of different functional components of the hearing system, providing simulated neural representations in the ascending auditory pathway up to the inferior colliculus. Its functioning was evaluated on the basis of human otoacoustic emissions \cite{Verhulst2012} and auditory evoked potentials \cite{Verhulst2018a,Verhulst2015}. The block diagram of the model is shown in Fig. \ref{fig:diagram}. The first processing stage of the model is the outer- and middle-ear models. The outer ear model \cite{Pralong1996} simulates the presentation of sounds through circumaural headphones. If the outer ear is omitted, then the sounds are assumed to be presented via in-ear earphones. After the outer- and middle-ear filters (Stage 1), the signal is fed into a transmission-line cochlear filter bank (Stage 2) where basilar membrane vibrations are simulated at 401 equidistant cochlear locations. A non-linear compression is applied to the cochlear vibrations, followed by a conversion into inner-hair-cell receptor potentials (Stage 3). These potentials are used to generate spike-rate patterns at the level of the auditory nerve (AN, Stage 4), which are related to fibres that have high- (HSR, 70 spikes/s), medium- (MSR, 10 spikes/s), and low-spontaneous rates (LSR, 1 spikes/s). These patterns are finally combined (13 HSR, 3 MSR, and 3 LSR fibres, i.e., 13-3-3) to form the input to a model of the spherical bushy cell in the cochlear nucleus (Stage 5) and inferior colliculus (Stage 6). The information across simulated characteristic frequencies available at the outputs of Stages 4, 5, and 6 offer a correlate of human brainstem responses recorded from scalp electrodes. The ``central processor stage'' (Stage 7) represents a way to integrate the simulated information in the ascending auditory pathway that is assumed to represent the way in which human listeners code and compare sounds among each other \cite{Osses2019b,Verhulst2018b}.

\vspace{-12pt}
\subsection*{Model update in v1.2}

In \textbf{model v1.2} two changes were introduced to the CN/IC model description (script \textsf{ic\_cn2018.py}, from the \href{https://github.com/HearingTechnology/Verhulstetal2018Model/releases}{GitHub} repository). One of the changes (correction of a bug) is relevant and affects the IC stage implementation  while the other is a minor change that was introduced to the lowpass filter design of alpha functions (\textbf{v1.2} using Eq.~(\ref{eq:coeffv12}), \textbf{v1.1} using Eq.~(\ref{eq:coeffv11}), see Section~\ref{sec:CN+IC}).\vspace{4pt}

A bug in the \textbf{v1.1} IC implementation resulted in a dominance of the excitatory neural responses while the effect of the inhibitory responses was nearly marginal. This situation was a consequence of using $1/\tau_{\textsubscript{exc}}^4$ and $1/\tau_{\textsubscript{inh}}^4$ as scaling constants for the excitatory and inhibitory signals, respectively, instead of $1/\tau_{\textsubscript{exc}}^2$ and $1/\tau_{\textsubscript{inh}}^2$ as shown later in Eq.~(\ref{eq:IC}). The excitation-dominated IC outputs were due to the much larger $1/\tau_{\textsubscript{exc}}^4$ value than $1/\tau_{\textsubscript{inh}}^4$. \vspace{4pt}

As a consequence of correcting this bug, the original M5 scaling factor (referred to as A\textsubscript{W-V} in the model paper \cite{Verhulst2018a}) had to be updated to maintain human-like ABR wave-V amplitudes. Because we needed to re-calibrate, we decided to adopt a more accurate procedure to calibrate the model constants to match the normative data by \citet{Picton2011-Ch08} than originally presented. This resulted in updated M1 (A\textsubscript{W-I} in the paper) and M3 (A\textsubscript{W-III} in the paper) scaling factors as well. The new calibration procedure is explained in Section~\ref{sec:const} and the obtained scaling factors are given in Table~\ref{tab:const}. \vspace{4pt}

\begin{table}[b!]
\caption{Scaling constants to yield ABR Waves I, III, and V to a realistic peak amplitude before (v1.1) and after (v1.2) correcting the bug and adopting a new calibration procedure. These constants are specified in the Python script \textsf{ic\_cn2018.py}, lines 5-7.}
\centering
\scalebox{0.9}{
\begin{tabular}{llccccc} \hline\hline
\multicolumn{2}{c}{Constant label} & \multicolumn{2}{c}{Value (dimensionless)} & \multicolumn{2}{c}{ABR amplitudes from} \\
Paper & Model& Model v1.2 (New) & \color{red}Model v1.1 & Model v1.2 & \color{red}Model v1.1 \\ \hline
A\textsubscript{W-I}   & M1 & $4.2767 \cdot 10^{-14}$ &  \color{red}$6.2755 \cdot 10^{-14}$ & 0.15 [$\mu$V$_{p}$]* & \color{red}0.30 [$\mu$V$_{pp}$] \\
A\textsubscript{W-III} & M3 & $5.1435 \cdot 10^{-14}$ &  \color{red}$7.2161 \cdot 10^{-14}$ & 0.17 [$\mu$V$_{p}$]* &\color{red} 0.34 [$\mu$V$_{pp}$] \\
A\textsubscript{W-V}   & M5 & $13.3093 \cdot 10^{-14}$ &  \color{red}$3.5200 \cdot 10^{-20}$& 0.61 [$\mu$V$_{pp}$] &\color{red} 0.61 [$\mu$V$_{pp}$] \\ \hline\hline 
\multicolumn{7}{l}{*~Picton's normative data \cite{Picton2011-Ch08} are reported to be 0.30, 0.34, 0.61 [$\mu$V] peak-to-peak for waves I, III, and}\\
\multicolumn{7}{l}{\hphantom{*~}V, respectively. We assumed that the corresponding baseline-to-peak voltages are approximately half}\\
\multicolumn{7}{l}{\hphantom{*~}of these values.}\\
\end{tabular}
}
\label{tab:const}
\end{table}

\newpage
\vspace{-20pt}
\section{CN and IC model implementation within model v1.2}
\label{sec:CN+IC}

\subsection{Analytical formulation}
\vspace{-6pt}
\textbf{(Same formulation as in model v1.1 \cite{Verhulst2018a})}\\

As stated in \cite{Nelson2004}, ventral CN postsynaptic potentials can be approximated by using alpha functions of the form $P(t)=t\cdot\exp{(-t/\tau)}$. In our model these functions are of the form $P(t)=t/\tau^2 \cdot\exp{(-t/\tau)}$. The area under the impulse response of the alpha functions is equal to 1. Given that alpha functions act as lowpass filters in the frequency domain, unit energy in the time domain corresponds to a 0-dB gain in the passband of the filter.\\

The originally described analytical CN model equation is \cite[][Eq. 17]{Verhulst2018a} (refer to the paper for a detailed description of each of the equation parameters):

\vspace{-12pt}
\begin{equation}
	r\textsubscript{CN}(t) = A\textsubscript{CN} \cdot \left[ \frac{t}{\tau_{\textsubscript{exc}}^2} \cdot \exp{\left(-\frac{t}{\tau_{\textsubscript{exc}}}\right)*r\textsubscript{AN}(t)} - S\textsubscript{CN,inh} \cdot \frac{t}{\tau_{\textsubscript{inh}}^2} \cdot \exp{\left(-\frac{t}{\tau_{\textsubscript{inh}}}\right)*r\textsubscript{AN}(t-D\textsubscript{CN})} \right]
\end{equation}

The IC model equation is:

\vspace{-8pt}
\begin{equation}
	r\textsubscript{IC}(t) = A\textsubscript{IC} \cdot \left[ \frac{t}{\tau_{\textsubscript{exc}}^2} \cdot \exp{\left(-\frac{t}{\tau_{\textsubscript{exc}}}\right)*r\textsubscript{IC}(t)} - S\textsubscript{IC,inh} \cdot \frac{t}{\tau_{\textsubscript{inh}}^2} \cdot \exp{\left(-\frac{t}{\tau_{\textsubscript{inh}}}\right)*r\textsubscript{IC}(t-D\textsubscript{IC})} \right]
	\label{eq:IC}
\end{equation}

Next, we provide a detailed analysis of how the function  $P(t)$ can be implemented in the discrete domain.

\subsection{Conversion to discrete domain}
\vspace{-6pt}
\textbf{(Minor change with respect to model v1.1)}\\

In our model (both, v1.1 and v1.2) alpha functions are implemented as IIR filters in the digital domain, with filter coefficients which are obtained from the corresponding $z$-transfer function $H(z^{-1})$. The continuous time domain function is $P(t)=t/\tau^2 \cdot\exp{(-t/\tau)}$. The frequency-domain implementation, from which $H(z^{-1})$ can be obtained, is given by the Laplace transform $H(s)$. For $P(t)=K \cdot t^n \cdot \exp{\left(a\cdot t\right)}$, with constants $K$ (here $K=1/\tau^2$), $a$ (here $a=-1/\tau$), and $n$ (here $n$=1):

\vspace{-8pt}
\begin{equation}
	H(s)=\frac{K\cdot n!}{\left(s-a\right)^{n+1}}\mbox{\hspace{30pt}}\implies\mbox{\hspace{30pt}}H(s)=\frac{K}{\left(s-a\right)^2}
\end{equation}

where $s=j\omega$ and $\omega$ is the natural frequency. To convert the Laplace transform into digital domain, a bilinear transformation can be applied (replacement of each $s$ by $\frac{2}{T} \cdot \left( \frac{1-z^{-1}}{1+z^{-1}}\right)$, with $T=1/f_s$,  \cite[e.g.,][Chapter~7]{Oppenheim1999}). The transfer function of the alpha function in the discrete domain thus becomes:

\vspace{-8pt}
\begin{equation}
	H(z^{-1})=\frac{K}{\left(\frac{2}{T} \cdot \left( \frac{1-z^{-1}}{1+z^{-1}}\right)-a\right)^2}
\end{equation}
\vspace{-8pt}

The coefficients of the corresponding IIR filter can be obtained by developing this expression: 

\vspace{-8pt}
\begin{equation}
	H(z^{-1})=\color{blue}\frac{K \cdot T^2}{4} \cdot \color{red}\frac{1}{\left(1-T\cdot a/2\right)^2} \color{black} \cdot \frac{1+2 z^{-1}+z^{-2}}{1-2\cdot   \left(\frac{1+T\cdot a/2}{1-T\cdot a/2}\right) z^{-1}+\left(\frac{1+T\cdot a/2}{1-T\cdot a/2}\right)^2 z^{-2}} \nonumber
\end{equation}

\vspace{-8pt}
Replacing $K$ by $1/\tau^2$, $T$ by $1/f_s$ and $a$ by $-1/\tau$ and reordering yields:

\vspace{-8pt}
\begin{equation}
	H(z^{-1})=\color{blue} \frac{1}{(2 \cdot f_s \cdot \tau)^2} \cdot \color{red}\frac{(2\cdot f_s\cdot\tau)^2}{\left(2\cdot f_s\cdot\tau+1\right)^2}\color{black}\cdot \frac{1+2 z^{-1}+z^{-2}}{1-2\cdot\left(\frac{2 \cdot f_s \cdot \tau-1}{2 \cdot f_s \cdot \tau+1}\right) z^{-1}+\left(\frac{2 \cdot f_s \cdot \tau-1}{2 \cdot f_s \cdot \tau+1}\right)^2 z^{-2}}
\end{equation}

The coloured terms are scaling factors. The factor highlighted in red was not considered in the CN/IC implementation in \textbf{v1.1}. This factor is, however, almost unity. For instance, if $f_s=20$~kHz and $\tau=2$~ms, then $(2\cdot f_s\cdot\tau)^2/\left(2\cdot f_s\cdot\tau+1\right)^2=0.9755$, or $-0.22$~dB.\\

Hence, the transfer function of the alpha function as implemented in \textbf{model v1.2} is (after simplification):

\vspace{-4pt}
\begin{equation}
	H(z^{-1})=\color{blue}\frac{1}{\color{red} \left(2\cdot f_s\cdot\tau+1\right)^2}\color{black}\cdot \frac{1+2 z^{-1}+z^{-2}}{1-2\cdot\left(\frac{2 \cdot f_s \cdot \tau-1}{2 \cdot f_s \cdot \tau+1}\right) z^{-1}+\left(\frac{2 \cdot f_s \cdot \tau-1}{2 \cdot f_s \cdot \tau+1}\right)^2 z^{-2}}
\label{eq:coeffv12}
\end{equation}

\vspace{-8pt}
In comparison, the implementation in \textbf{model v1.1} was:

\vspace{-8pt}
\begin{equation}
	H(z^{-1})=\color{blue} \frac{1}{(2 \cdot f_s \cdot \tau)^2} \color{black}\cdot \frac{1+2 z^{-1}+z^{-2}}{1-2\cdot\left(\frac{2 \cdot f_s \cdot \tau-1}{2 \cdot f_s \cdot \tau+1}\right) z^{-1}+\left(\frac{2 \cdot f_s \cdot \tau-1}{2 \cdot f_s \cdot \tau+1}\right)^2 z^{-2}}
\label{eq:coeffv11}
\end{equation}

\vspace{-8pt}
The coefficients of the transfer function (and of the IIR filter) for model versions \textbf{v1.2} and \textbf{v1.1} are shown in Table~\ref{tab:coeff} and correspond to the constants in the numerator and denominator of Eqs.~\ref{eq:coeffv12} and \ref{eq:coeffv11}, respectively. For comparison, the IIR filter coefficients from the CN/IC implementation within the UR EAR toolbox v2.1\footnote{The toolbox is available at \href{https://www.urmc.rochester.edu/MediaLibraries/URMCMedia/labs/carney-lab/}{https://www.urmc.rochester.edu/MediaLibraries/URMCMedia/labs/carney-lab/} (Version 2.1 was downloaded on 2/03/2019).\vspace{-20pt}} (IC/CN model described in \cite{Nelson2004}) are also shown in the table. \vspace{-12pt}

\begin{table}[h!]
\caption{Coefficients of the transfer function (IIR filter) used within the CN and IC models.}
\scalebox{0.85}{
\begin{tabular}{cccccccccc} \hline\hline
Model   & \multicolumn{3}{c}{Numerator} & \multicolumn{3}{c}{Denominator} & & Scaling & Passband\\  
version & $b_0$ & $b_1$ & $b_2$ & $a_0$ & $a_1$ & $a_2$ &$m$ & factor C & gain G [dB]\\ \hline
v1.2 (Eq.~\ref{eq:coeffv12}) & 1 & 2 & 1 & 1 & $-2 \cdot m$& $m^2$ & $\frac{2\cdot f_s \cdot \tau-1}{2\cdot f_s \cdot \tau+1}$ & $\frac{1}{(2 \cdot f_s \cdot \tau+1)^2}$ & 0 \\
v1.1 and & \multirow{2}{*}{1} & \multirow{2}{*}{2} & \multirow{2}{*}{1} & \multirow{2}{*}{1} & \multirow{2}{*}{$-2 \cdot m$} & \multirow{2}{*}{$m^2$} & \multirow{2}{*}{$\frac{2\cdot f_s \cdot \tau-1}{2\cdot f_s \cdot \tau+1}$} & \multirow{2}{*}{$\frac{1}{(2 \cdot f_s \cdot \tau)^2}$} & 0.22 ($\tau=2$~ms)* \\
earlier (Eq.~\ref{eq:coeffv11}) &&&&&&&&  & 0.85 ($\tau=0.5$~ms)* \\ 
UR EAR & 0 & $m$ & 0 & 1 & $-2\cdot m$ & $m^2$ & $\exp{\left(-\frac{1}{f_s\cdot \tau}\right)}$ & $\frac{1}{f_s^2 \cdot \tau^2 \cdot \left[ 1-(\tau+1) \cdot \exp{(-1/\tau)}\right]}$ & 0 \\ \hline\hline
\multicolumn{10}{l}{\footnotesize{*Gains assessed using the default sampling frequency $f_s=20$~kHz that is used for simulated neural responses.}}
 \vspace{-10pt}
\end{tabular} 
}
\label{tab:coeff}
\end{table}

\subsection{Modulation transfer function (MTF) for AM tones}

\begin{figure}[b!]
	\centering
	\includegraphics[width=0.4\textwidth,trim=0 1.5cm 0 0,clip=true]{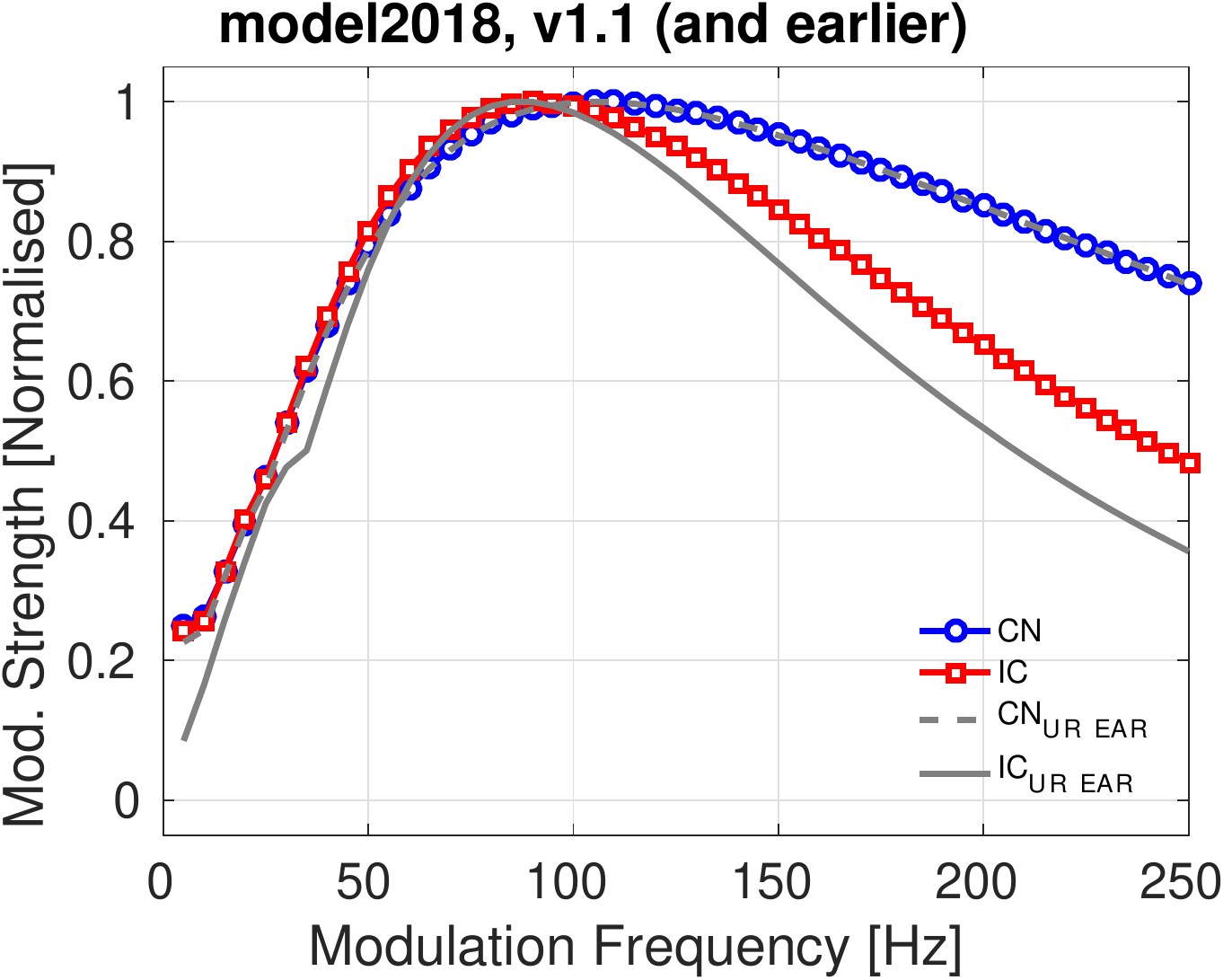}
	\includegraphics[width=0.4\textwidth,trim=0 1.5cm 0 0,clip=true]{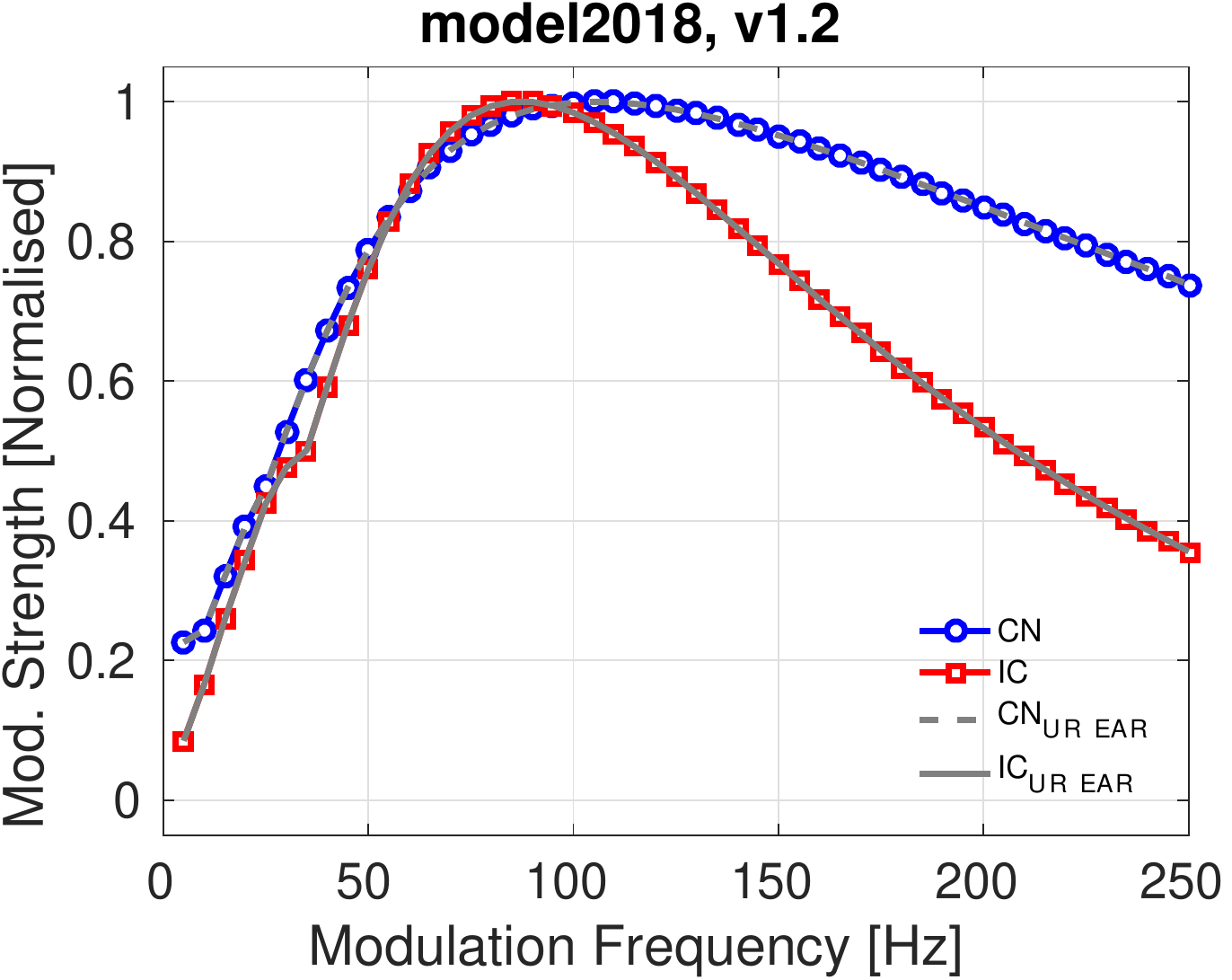}
	\includegraphics[width=0.4\textwidth,trim=0 0 0 0.7cm,clip=true]{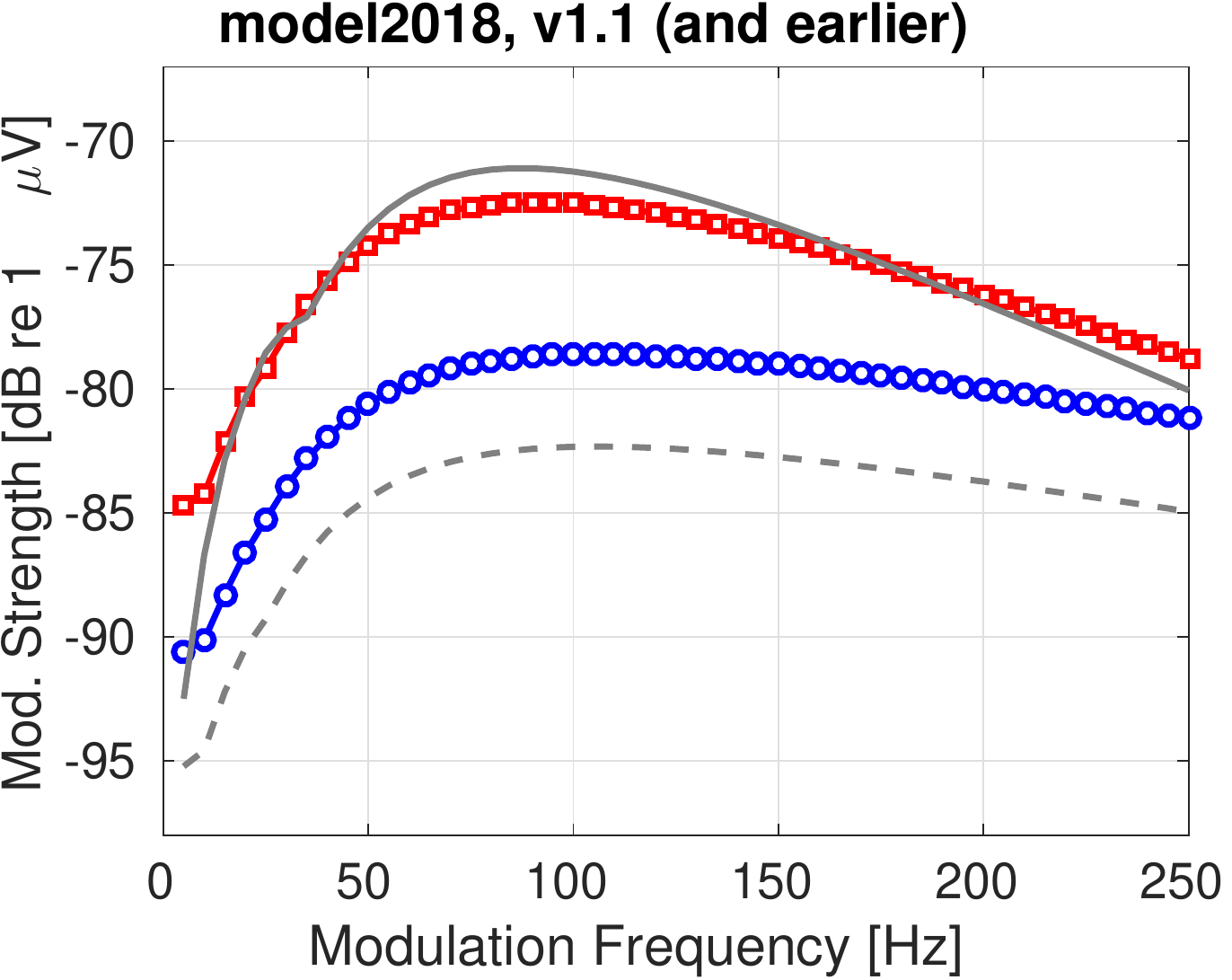} 
	\includegraphics[width=0.4\textwidth,trim=0 0 0 0.7cm,clip=true]{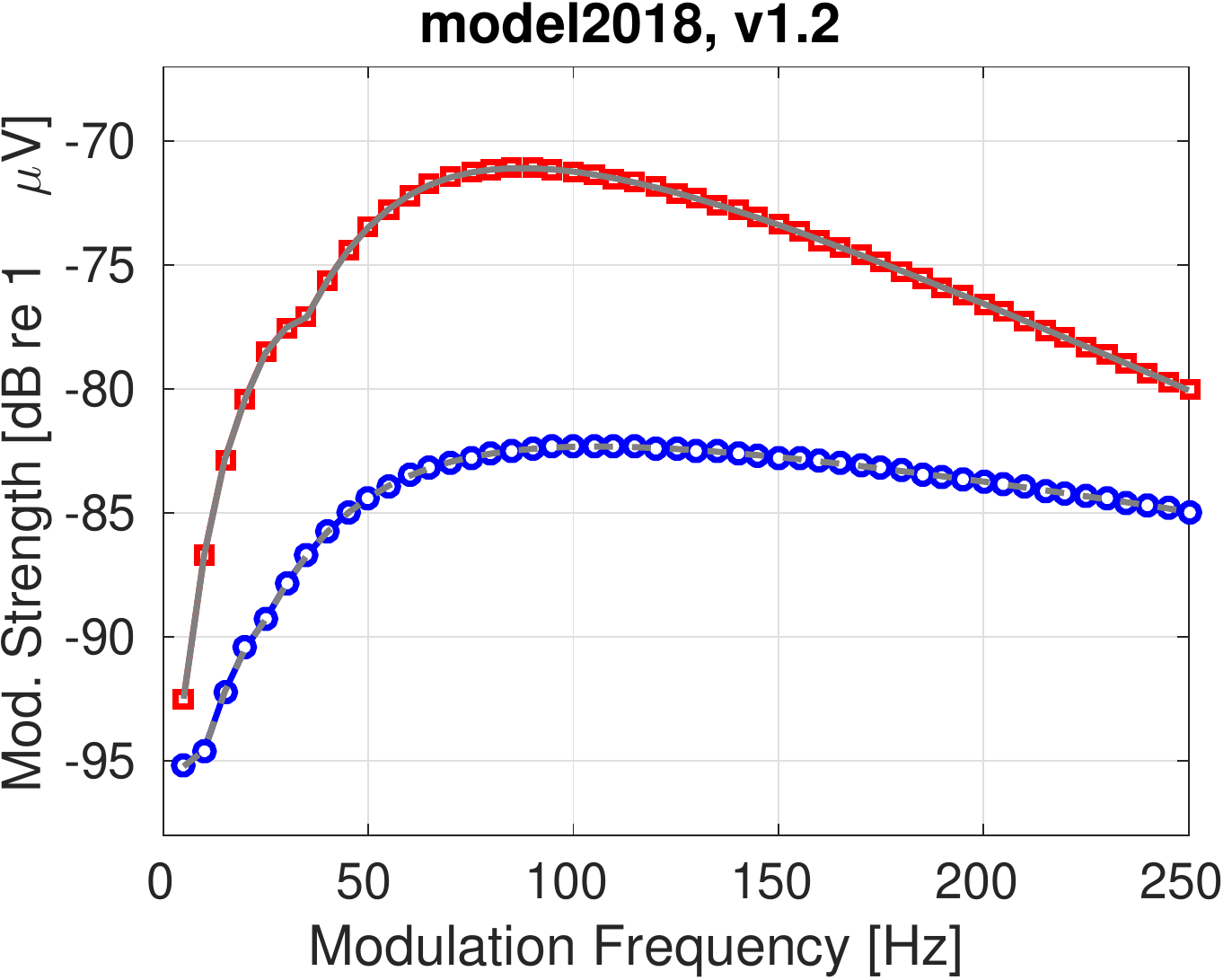}\\
	\vspace{-6pt}
	\caption{MTFs using model \textbf{v1.1} (left panels) and \textbf{v1.2} (right panels). The CN and IC outputs using model2018+UR EAR are indicated by the dashed and continuous grey lines, respectively.}
	\label{fig:mod-strength}
\end{figure}

To illustrate how the modifications to the CN/IC model description affect the filter tuning of the CN/IC models (\textbf{model v1.2}, \textbf{model v1.1}) the MTF for 4-kHz, 100\% modulated amplitude-modulated (AM) tones, was assessed. Our implementation is compared with the CN/IC implementation from Laurel Carney's group available within the UR EAR toolbox, which was used to process the AN responses obtained from our model. This model variant is labelled as \textbf{model2018+UR EAR}.\\

\textbf{Stimuli}: AM tones with carrier frequency f$_c=4$~kHz, duration of 200~ms, level of 70~dB SPL, gated with 5-ms up-down ramps were used. The tones were 100\% modulated with modulation rates of f\textsubscript{mod} between 5 and 250 Hz in steps of 5 Hz.\vspace{4pt}

\textbf{MTF assessment}: On-frequency MTFs for CN and IC outputs in channel 112 (i.e., at CF$=4013$~Hz) were assessed. The MTFs were obtained by reading the maximum amplitude of the corresponding 4000-point FFT, after DC removal. The DC removal was needed to obtain a maximum in the FFT response elicited by the sensitivity to stimulus modulations and not by the DC.\vspace{4pt}

\textbf{Results}: The obtained MTFs are shown in Fig. \ref{fig:mod-strength} for the different model implementations. The obtained on-frequency MTFs are shown in two forms: (1) As relative modulation strength (left panels), i.e., as simulated (maximum) voltages, normalised to the maximum modulation strength (at around f\textsubscript{mod}$=100$~Hz); (2) As maximum voltages in dB referenced to 1~$\mu V$. The bandpass characteristic of the CN and IC models is similar in the three evaluated models (model \textbf{v1.1}, \textbf{v1.2}, and \textbf{UR EAR}). The bandpass sensitivity (or modulation strength) is, however, slightly sharper for \textbf{v1.2} and \textbf{UR EAR} in comparison with \textbf{v1.1}. The MTFs between model \textbf{v1.2} and \textbf{model2018+UR EAR} show a sensitivity in the same magnitude range (superimposed grey and coloured curves in right panels).

\section{Model validation} 

To validate \textbf{model v1.2}, we simulated responses to the following stimuli:
\begin{itemize}
	\item Clicks of alternating polarity (for calibration of scaling factors in \textbf{v1.2}) 
	\item Clicks of positive polarity to reproduce Fig.~6 in \cite{Verhulst2018a}
	\item AM tones with carrier frequency of 4-kHz and 85\% modulation depth to reproduce Fig.~7 in \cite{Verhulst2018a}
\end{itemize}

All model simulations were performed in MATLAB (tested in various MATLAB releases between R2012b and R2018a, on Windows and Unix-based machines) and Python (v2.7) using the following configuration:

\begin{lstlisting}[caption=Configuration in model versions \textbf{v1.1} and \textbf{v1.2}. The basilar-membrane pole admittances used as input to the model (\textsf{sheraP}) corresponded to a normal hearing cochlear gain profile (``Flat00'') unless otherwise stated.\vspace{6pt}]
flag_cf='abr'; % 401 CF channels
flag_proc='cfvihmlbw';
non_linear_type='vel'; 
numH=13; numM=3; numL=3; % 19 neurones
dir_out=[pwd(),filesep]; % directory were variables will be stored
output = model2018(insig,fs,flag_cf,1,flag_proc,1,sheraP,0.05,non_linear_type,numH,numM,numL,1,dir_out);
\end{lstlisting} \vspace{-12pt}
\small{\hphantom{*}*See also \textsf{Example\_Simulation.m} from the \href{https://github.com/HearingTechnology/Verhulstetal2018Model/releases}{GitHub} repository.}

\vspace{-8pt}
\subsection{Clicks: Calibration of the scaling factors}
\label{sec:const}

\textbf{Calibration stimuli}: A click train with repetition rate of 11.1 Hz and duration of 6~s (e.g., 66 clicks) was generated. The clicks had alternating polarity (starting with positive) and each click had a duration of $80 \mu$s. The clicks with positive and negative polarities had amplitudes of $A$ and $-A$, respectively, which were matched to have a level of 100 dB peSPL, that is, having a baseline-to-peak amplitude equal to the peak-to-peak amplitude of a sinusoid of 100 dB SPL. The resulting clicks had an amplitude $A$ of $5.66$~[Pa].\vspace{4pt}

\begin{figure}[!h]
	\centering
	\includegraphics[width=0.85\textwidth,clip=true,trim=0 0 0 0.5cm]{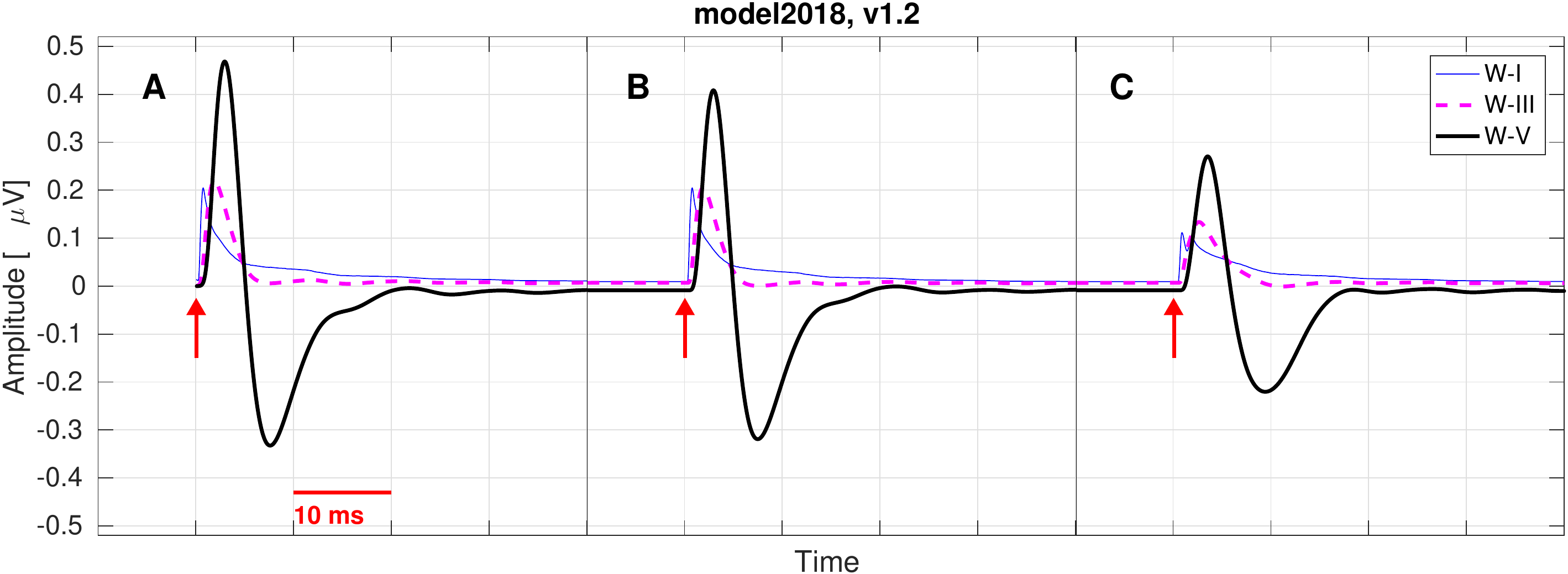}
	\vspace{-10pt}
	\caption{Clicks \#1 (\textbf{A}, positive polarity), \#59 (\textbf{B}, positive polarity) and \#60 (\textbf{C}, negative polarity) with an interstimulus time of about 90~ms (11.1~Hz presentation rate, not schematised in this figure). Clicks \#59 and 60 (used for the model calibration) have baseline-to-peak amplitudes of 0.20 and 0.10$\mu$V for W-I (average of 0.15$\mu$V, in blue), 0.21 and 0.13$\mu$V for W-III (average of 0.17$\mu$V, in magenta), and peak-to-peak amplitudes of 0.73 and 0.49$\mu$V for W-V (average of 0.61$\mu$V, in black), respectively. As a consequence of the calibration process, the average peak amplitudes of the clicks match the amplitudes of the normative data \cite{Picton2011-Ch08} (see Table~\ref{tab:const}). The red arrows indicate the click onset time.}
	\label{fig:clicks}
\end{figure}

\textbf{Simulated of ABR amplitudes}: As shown in Fig.~\ref{fig:clicks}, broadband W-I and W-III are mainly excitatory responses while the simulated W-V has a combined excitatory/inhibitory pattern (with both positive and negative voltage deflections). To compensate for the lack of inhibition in simulated W-I and W-III, the peak-to-peak W-I and W-III amplitudes are assumed to be double as large as the baseline-to-peak amplitudes and this implies that the through-voltage is assumed to have the same amplitude (with opposite sign) than the corresponding peak value. For this reason, \textbf{model 2018 v1.2} is suitable to simulate baseline-to-peak W-I and W-III voltages and a peak-to-peak W-V voltage. \textbf{In order to express all simulated amplitudes that are obtained with model v1.2 in peak-to-peak values, W-I and W-III [V$_p$] have to be multiplied by a factor of 2.} This scaling factor was not needed for estimated amplitudes using \textbf{model v1.1} because in that model version all W-I, W-III, and W-V showed excitatory responses such that baseline-to-peak and peak-to-peak amplitudes corresponded with each other.\vspace{4pt}

\textbf{Calibration procedure}: The scaling factors M1, M3, and M5 were adjusted to provide calibrated amplitudes of simulated waveforms (wave-I, -III, and -V, respectively) in volts (V) according to the normative data published in \cite[][Table 8--1]{Picton2011-Ch08} for clicks at a level of 70 dB nHL (the approximation used in the calibration procedure is that 100 dB peSPL$\approx$70 dB nHL). In addition to the multiplication by a factor of 2 to estimate V$_{pp}$ values for W-I and W-III when using \textbf{model v1.2} with respect to \textbf{v1.1}, the calibration procedure in \textbf{v1.2} is based on later clicks along the click train (illustrated here using clicks \#59,60), while previous model versions were calibrated using the response of the first click only (click~\#1).\vspace{4pt}

\textbf{Results}: The model response to the click train using the obtained scaling factors (Table~\ref{tab:const}) is illustrated in Fig.~\ref{fig:clicks} for clicks \#1, 59, and 60. Clicks \#59 and 60 have average W-I, W-III, and W-V amplitudes of 0.15 [$\mu$V$_p$], 0.17 [$\mu$V$_p$], and 0.61 [$\mu$V$_{pp}$], which represent estimated peak-to-peak values (two times the simulated W-I and W-III amplitudes) of 0.30, 0.34, and 0.61 [$\mu$V$_{pp}$]. 

\subsection{Clicks: ABR wave-I and wave-V characteristics}

\begin{figure}[!b]
	\centering
	\subfigure[\textbf{Simulated ABR W-V latencies} as a function of click level for models using a normal-hearing profile (NH: Flat00, 13-3-3, green line), a sloping high-frequency hearing-loss profile (HI: Slope35, 13-3-3, purple line), and a synaptopathy profile has contributions from only HSR neurones (HSR: Flat00, 13-0-0, red dashed line). The simulated latencies were incremented by 3.5~ms to compare the simulations with the corresponding reference data (which are not shown here), as it was also performed in the model~paper. These panels are similar to Fig.6A from \cite{Verhulst2018a}.\vspace{-4pt}]
	{
	\includegraphics[height=0.20\textheight                           ]{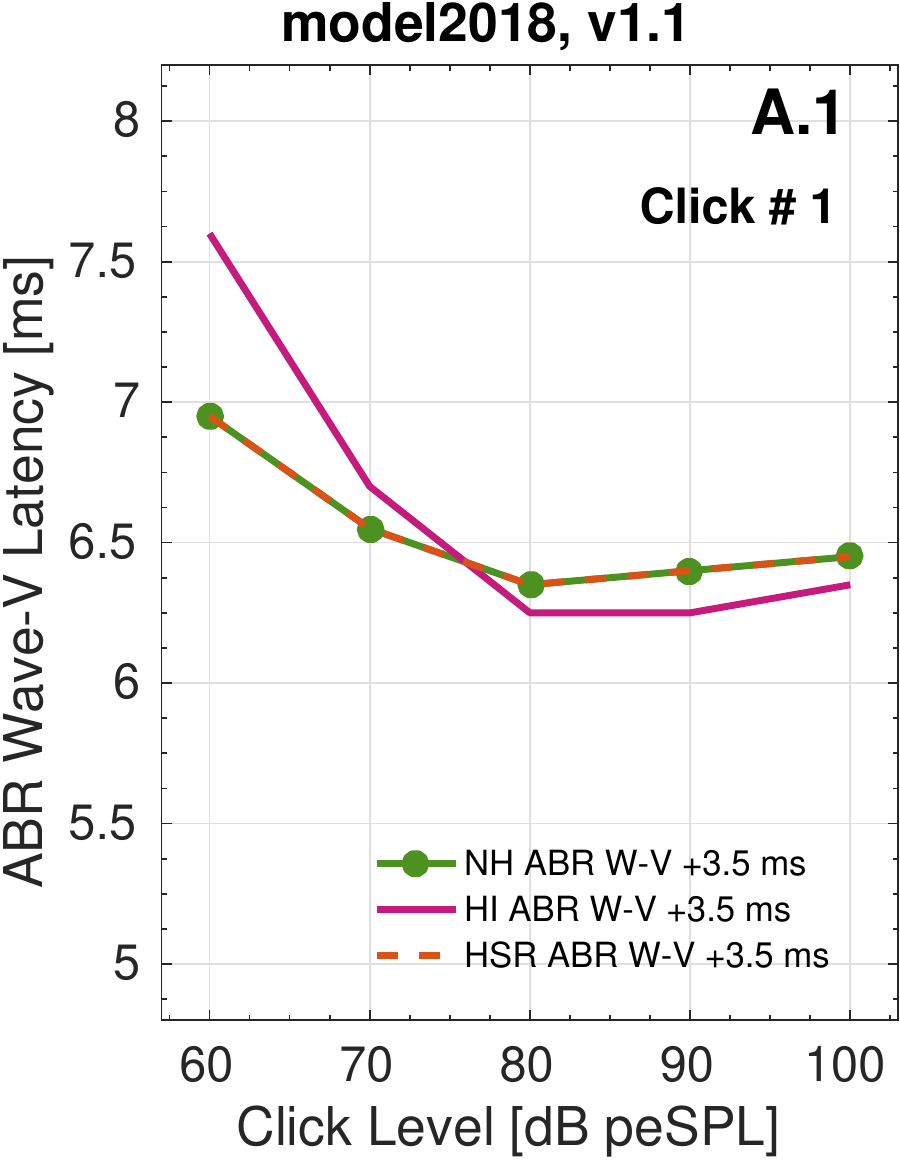}
	\includegraphics[height=0.20\textheight,trim=1.41cm 0 0 0,clip=true]{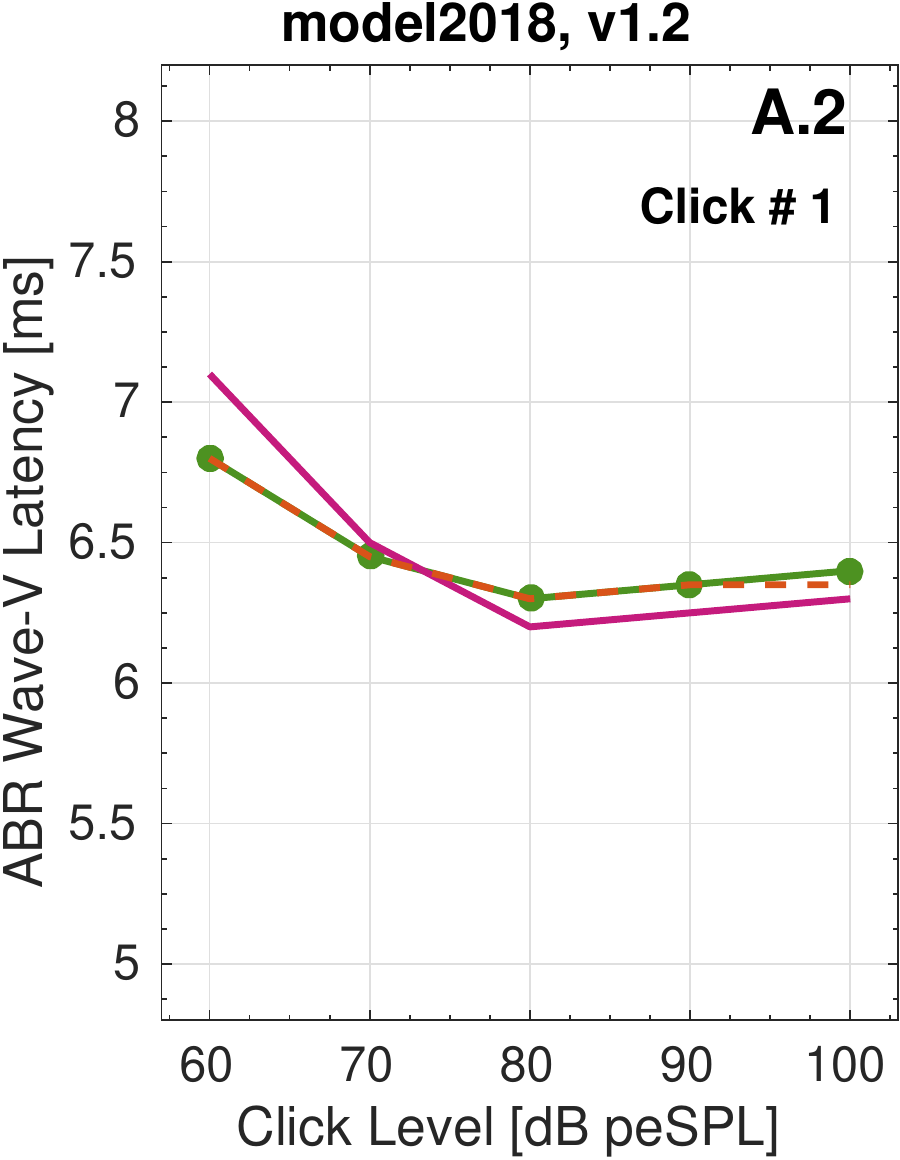}
	\includegraphics[height=0.20\textheight,trim=1.41cm 0 0 0,clip=true]{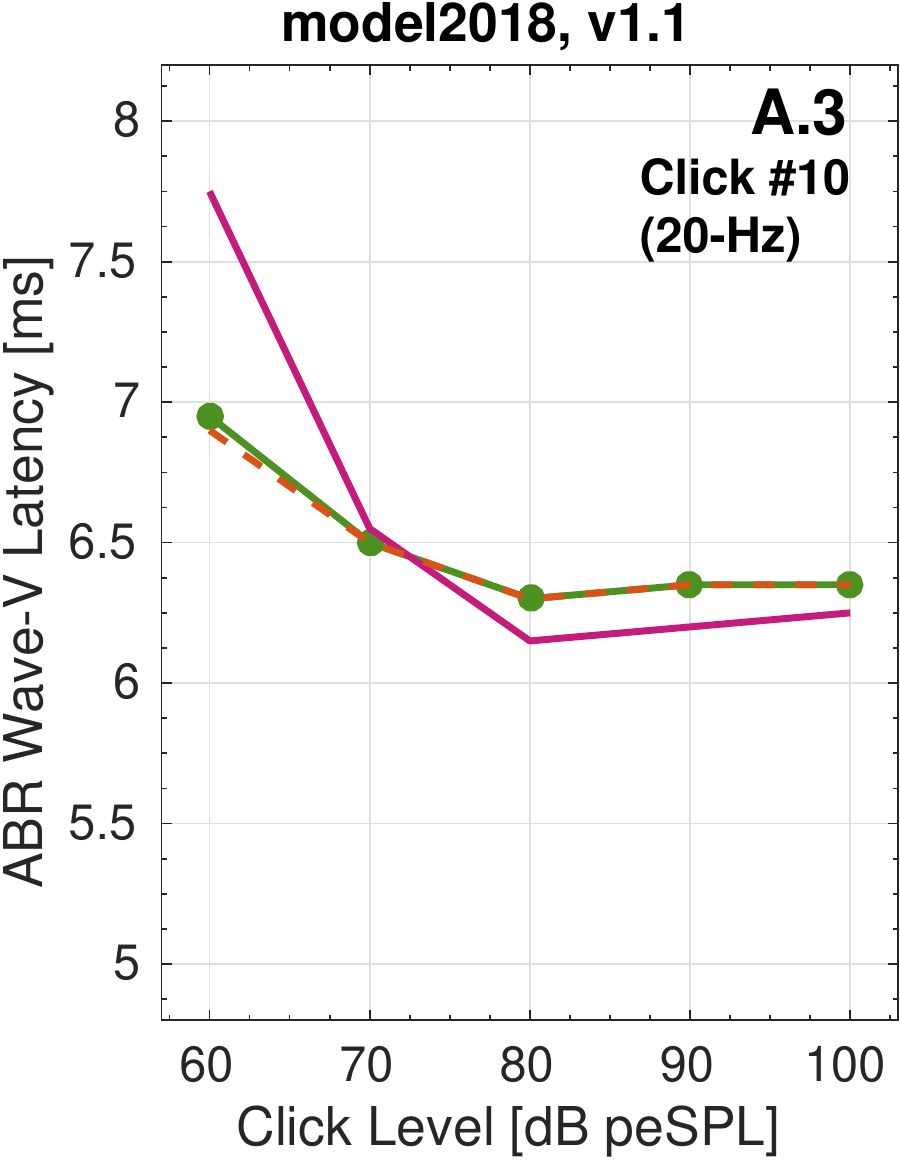}
	\includegraphics[height=0.20\textheight,trim=1.41cm 0 0 0,clip=true]{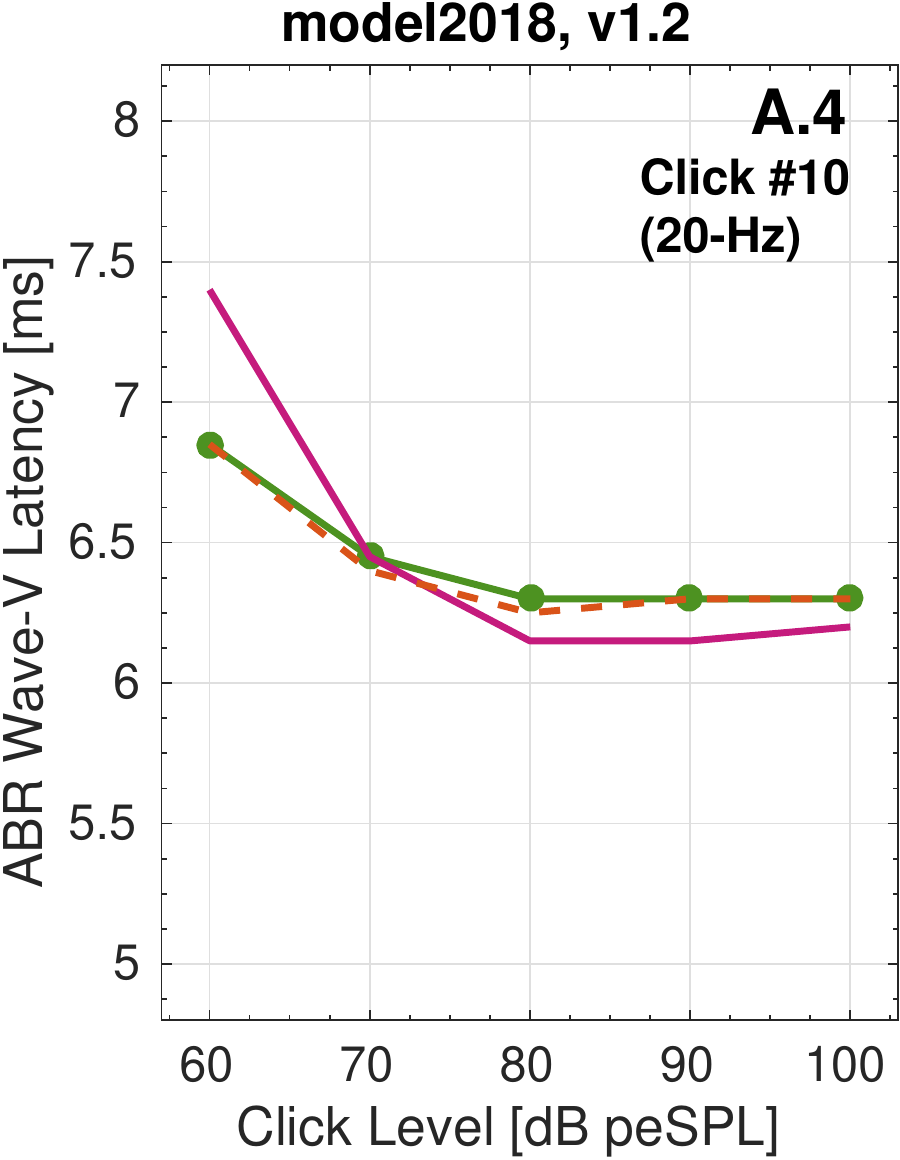}
	}
	\subfigure[Same simulated latencies as reported in panel A, but here only the results for the NH profile are shown, without the 3.5-ms increment. These panels are similar to Fig.6B from \cite{Verhulst2018a}.\vspace{-4pt}]
	{
	\includegraphics[height=0.19\textheight,trim=0 0 0 0.6cm,clip=true]{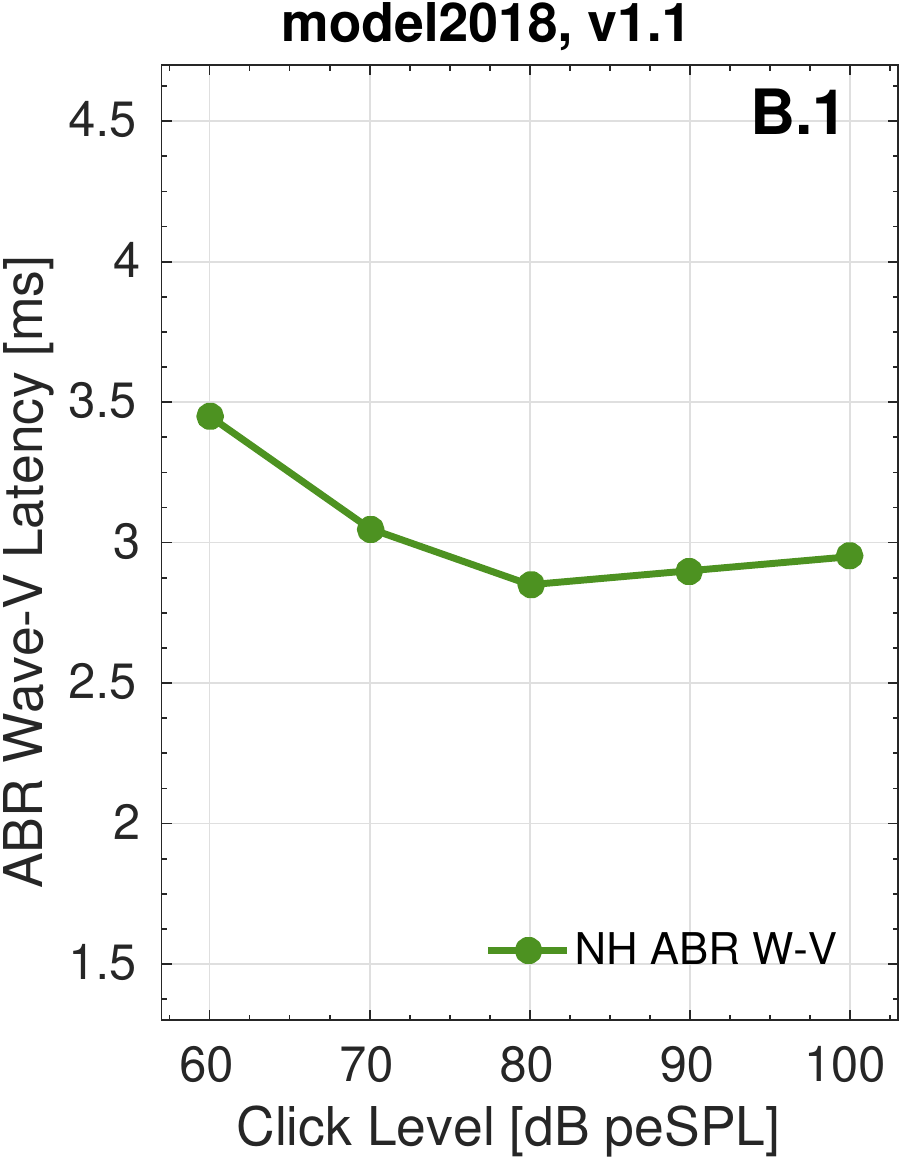}
	\includegraphics[height=0.19\textheight,trim=1.4cm 0 0 .6cm,clip=true]{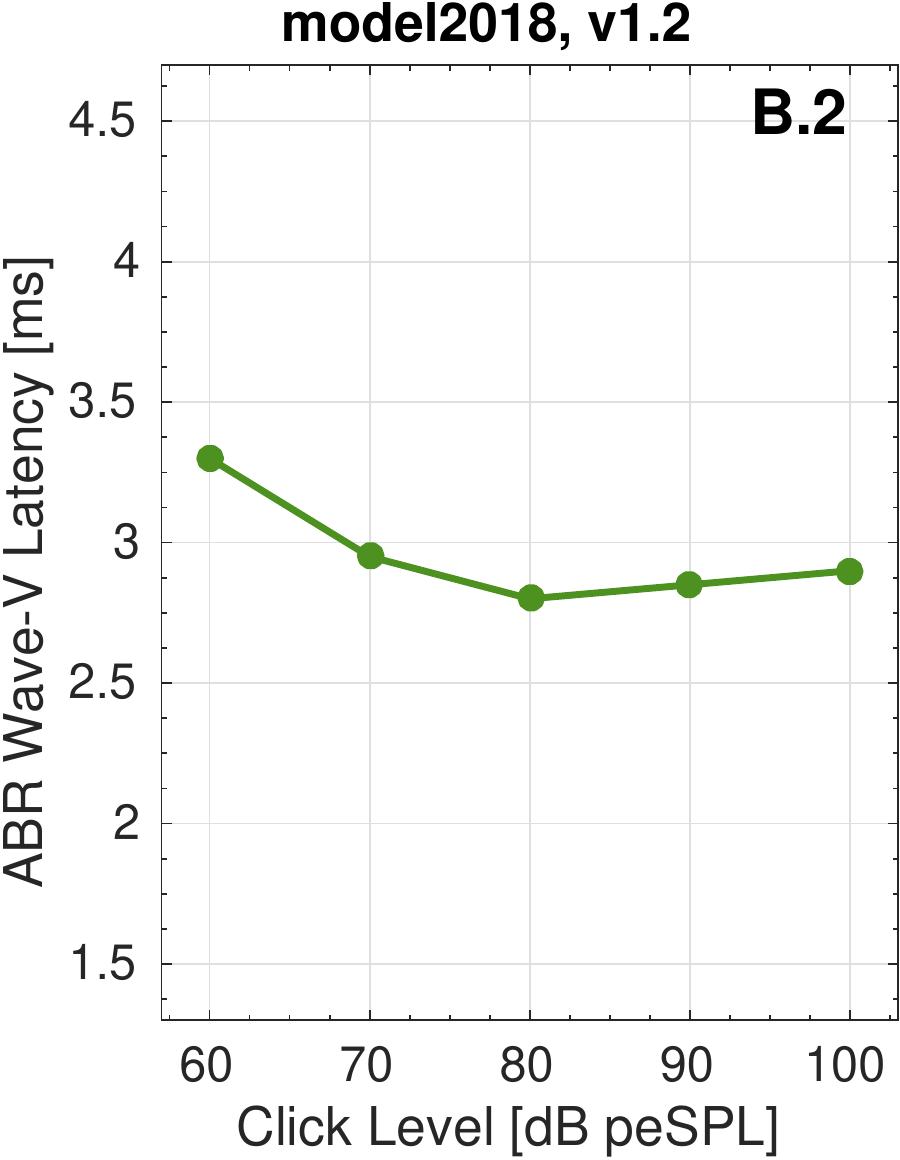}
	\includegraphics[height=0.19\textheight,trim=1.4cm 0 0 .6cm,clip=true]{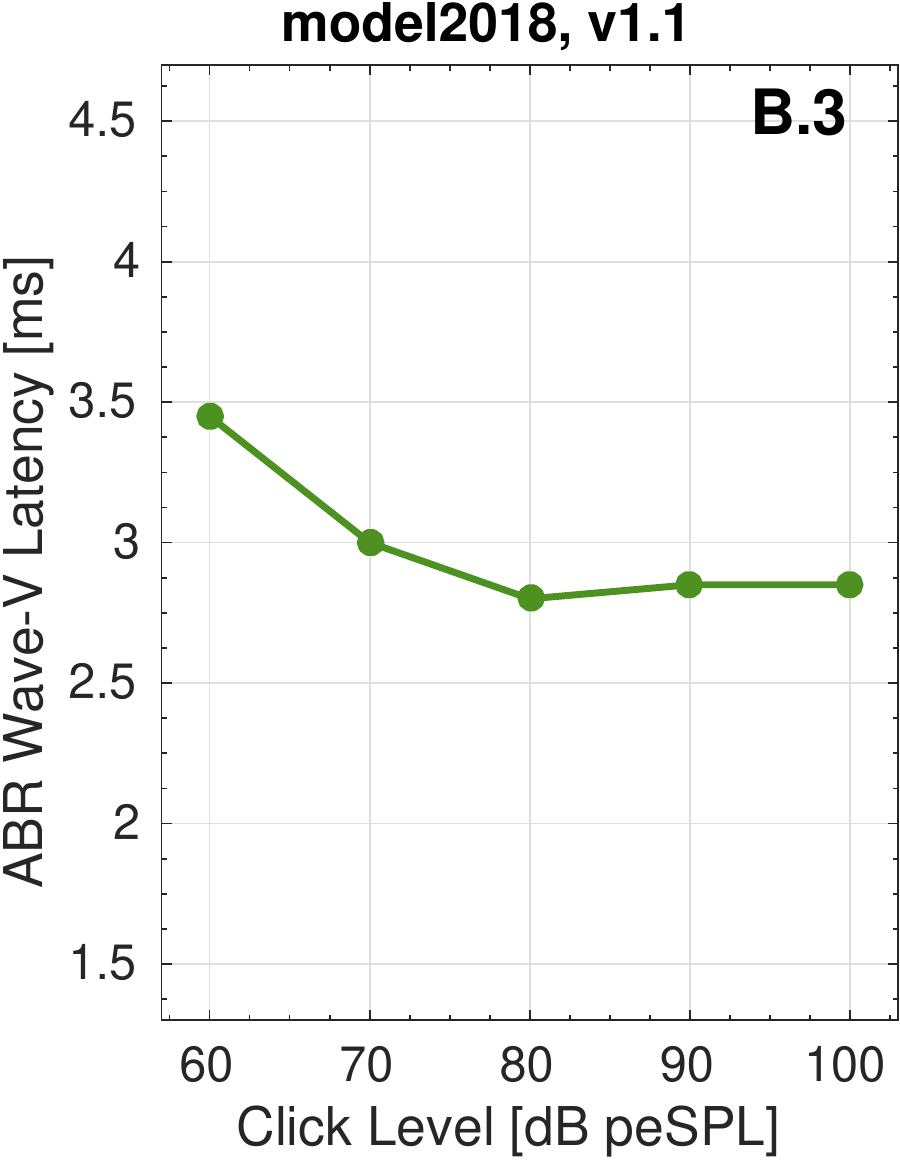}
	\includegraphics[height=0.19\textheight,trim=1.4cm 0 0 .6cm,clip=true]{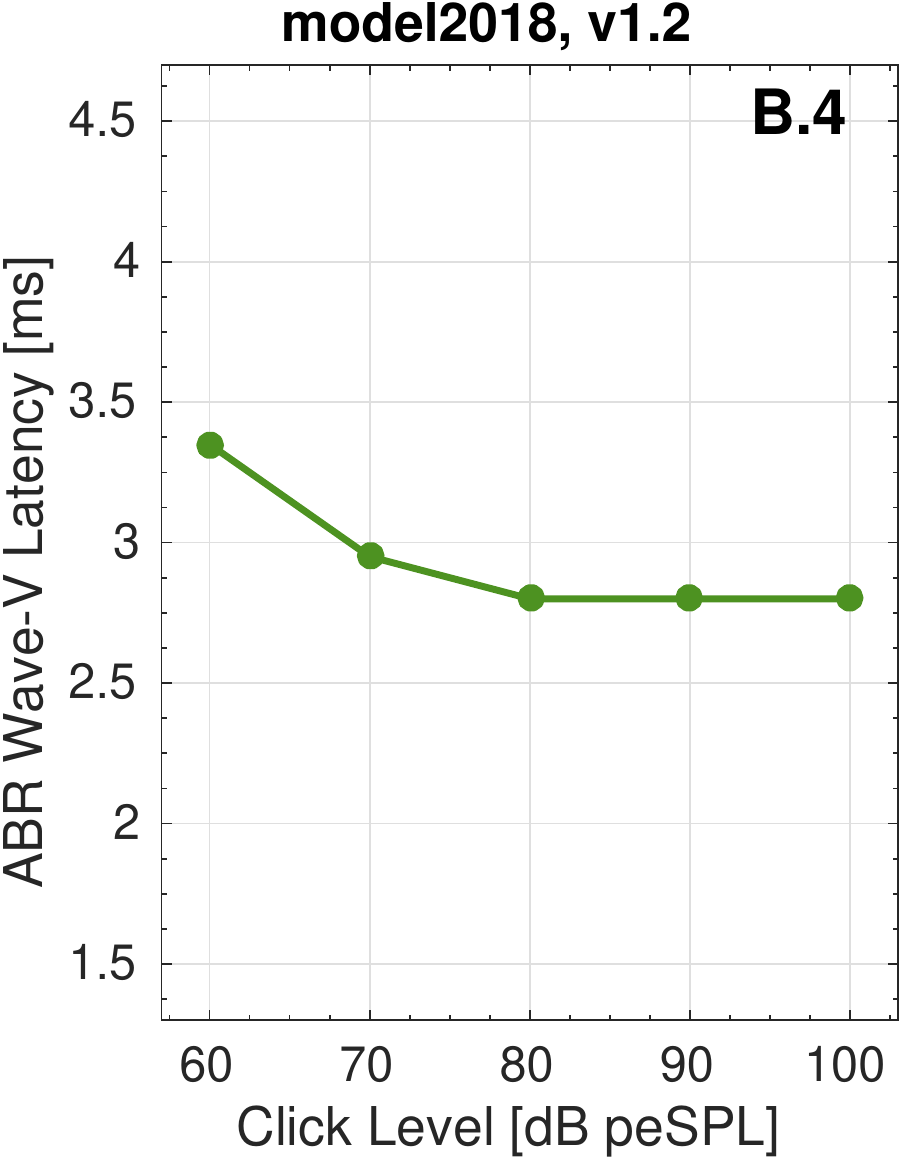}
	}
	\subfigure[\textbf{Simulated ABR W-I and W-V amplitudes} as a function of click levels for models using the NH profile (green lines), the HI profile (purple lines), and a selective synaptopathy profile HSR (red dashed lines). In general, cochlear gain loss (HI profile) steepened the ABR amplitude growth curve and synaptopathy (HSR profile) resulted in shallower curve slopes compared to the NH profile. These panels are similar to Fig. 6C from \cite{Verhulst2018a}.]
	{
	\includegraphics[height=0.19\textheight,trim=0 0 0 0.6cm,clip=true]{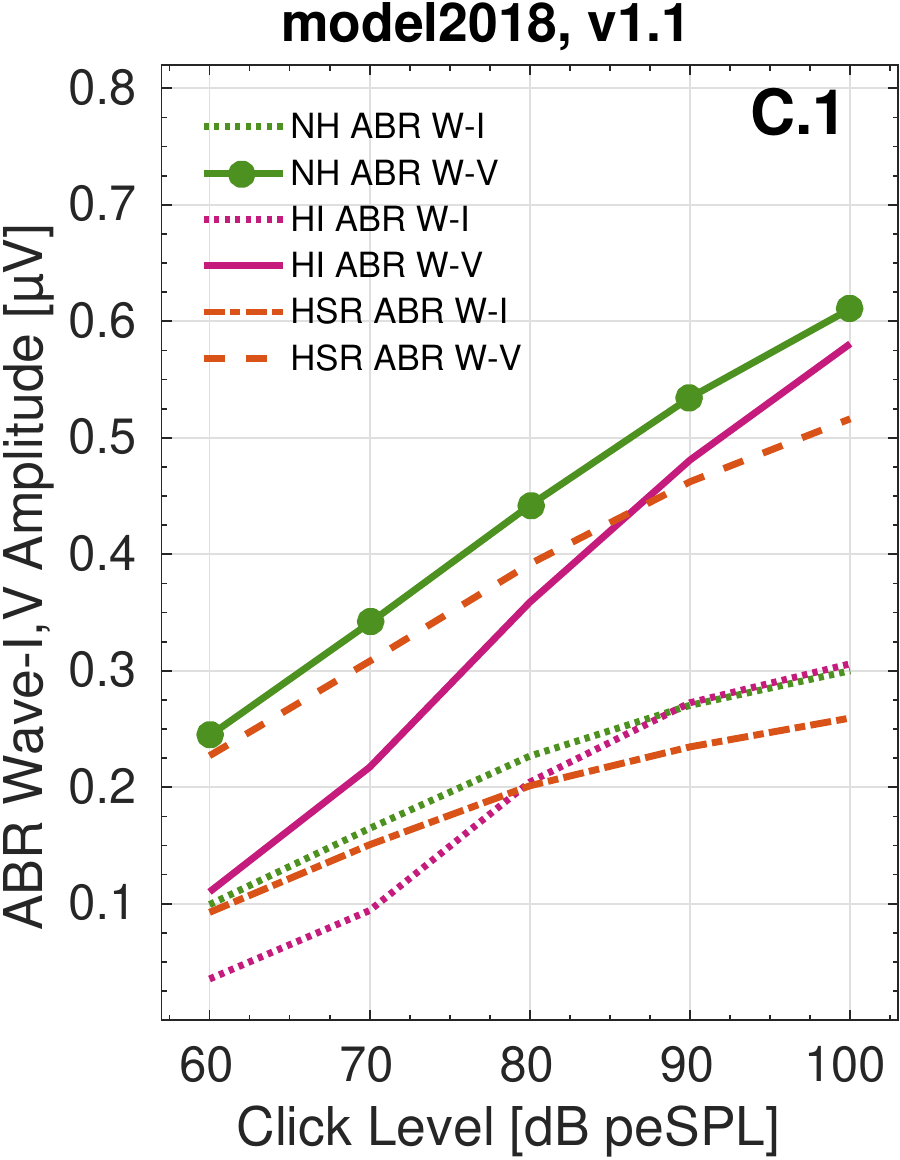}
	\includegraphics[height=0.19\textheight,trim=1.4cm 0 0 0.6cm,clip=true]{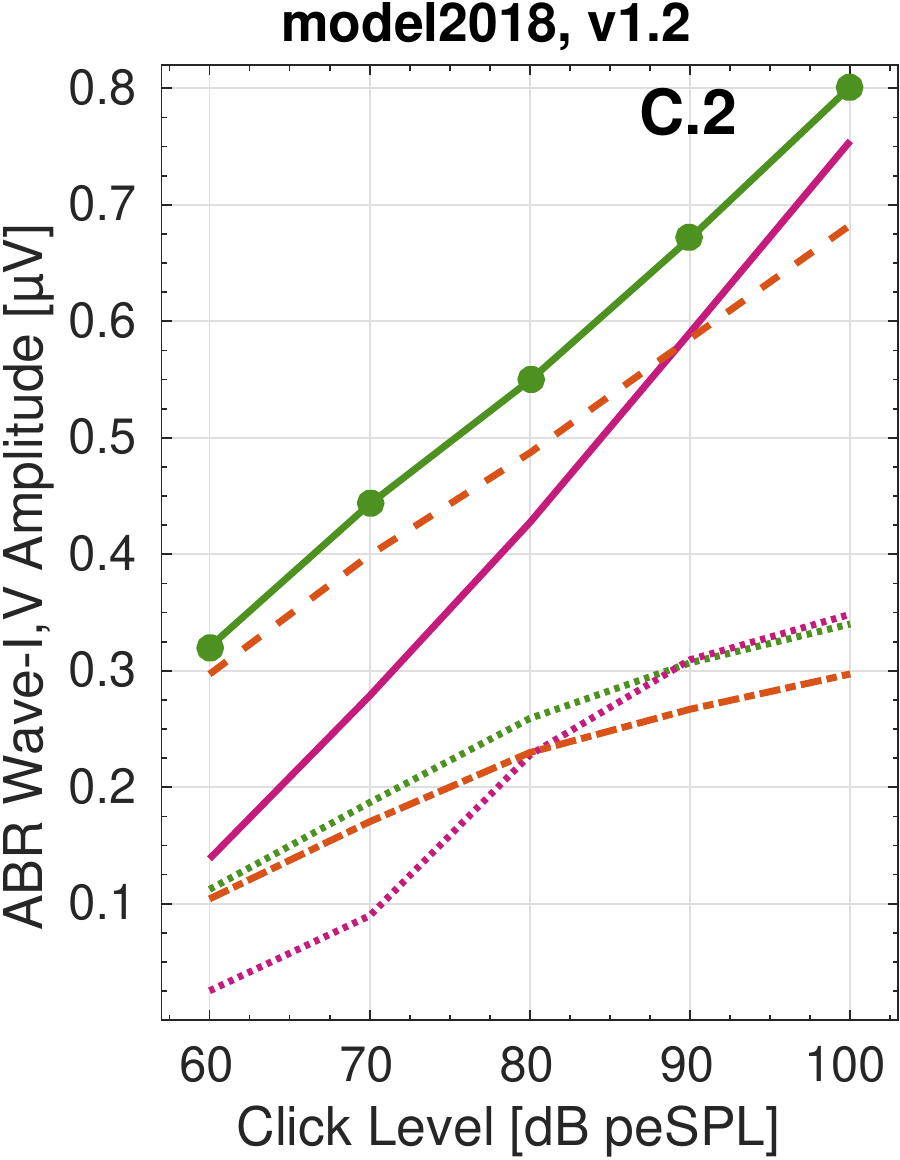}	
	\includegraphics[height=0.19\textheight,trim=1.4cm 0 0 0.6cm,clip=true]{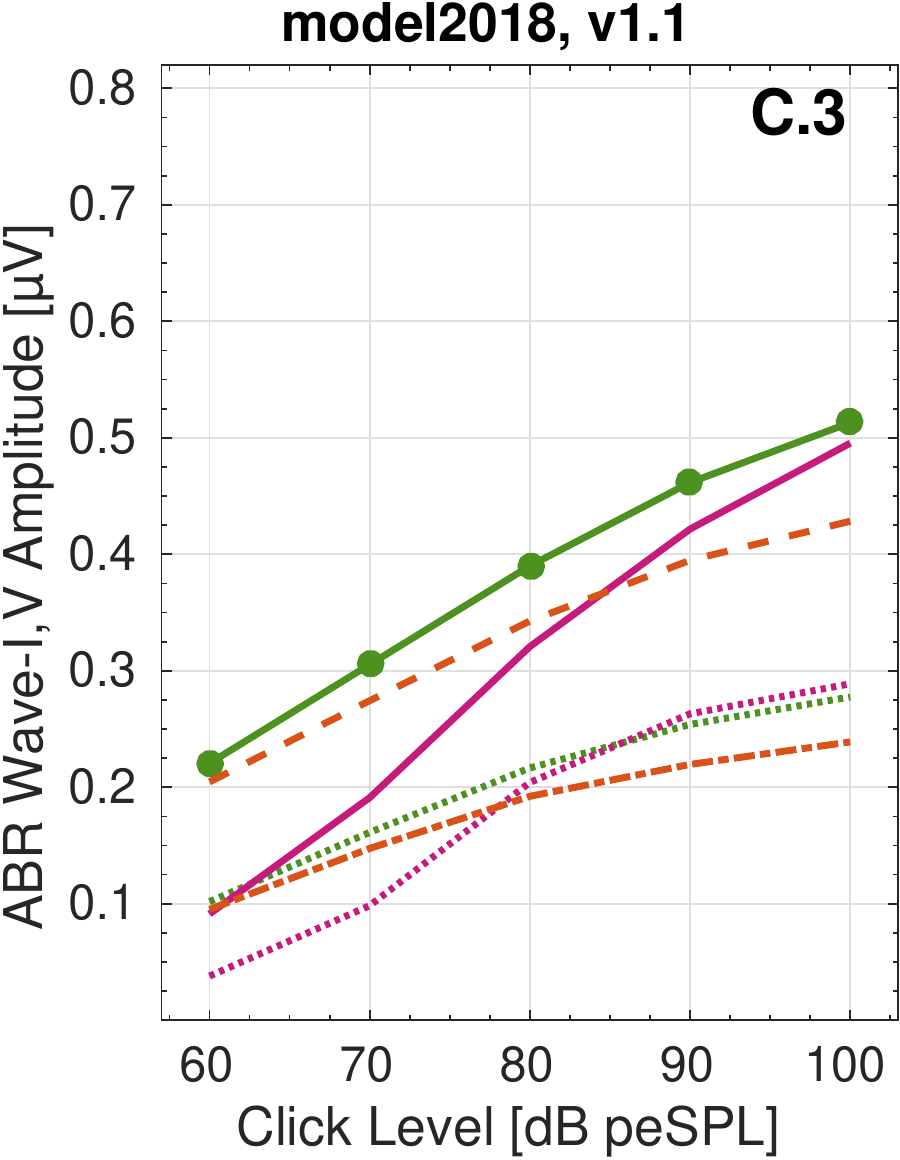}	
	\includegraphics[height=0.19\textheight,trim=1.4cm 0 0 0.6cm,clip=true]{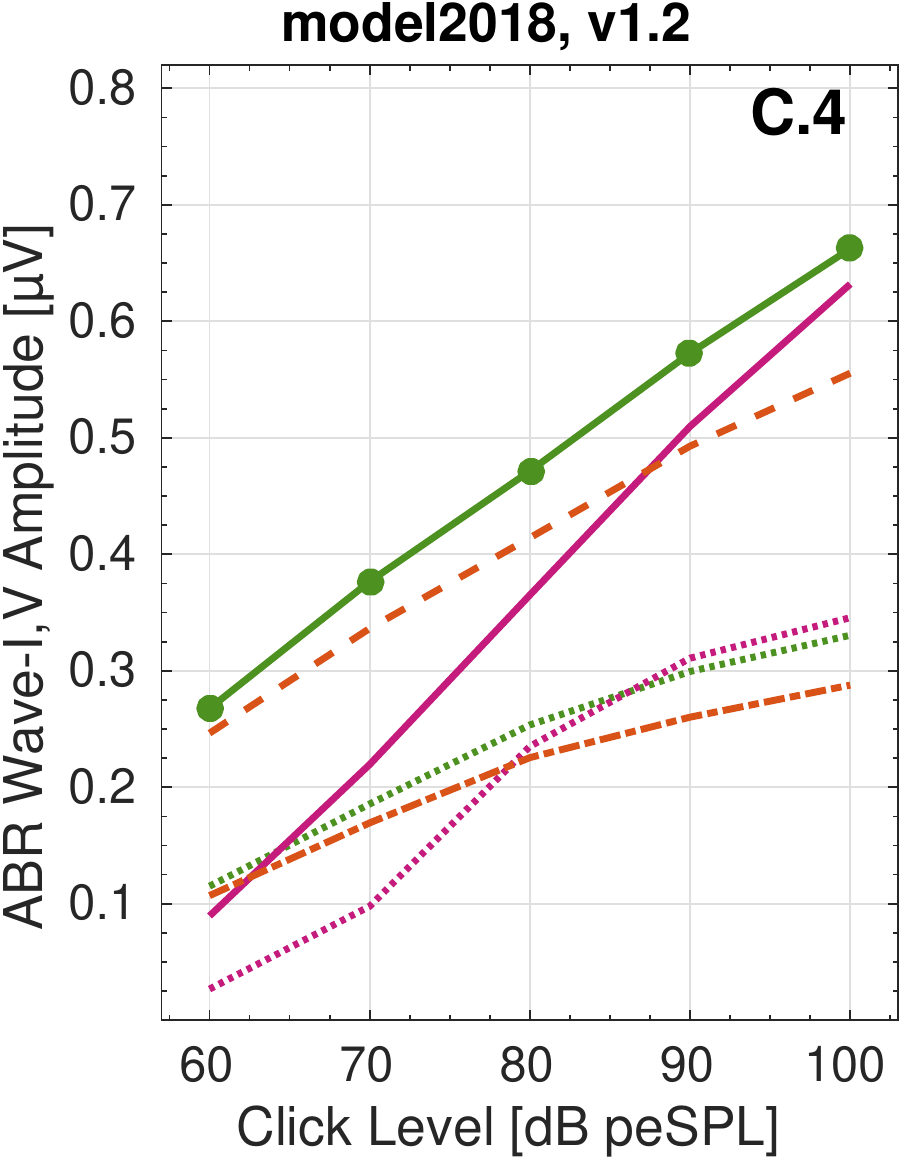}	
	}
	\vspace{-10pt}
	\caption{Simulations using \textbf{model v1.1} (panel columns 1 and 3) and \textbf{model v1.2} (panel columns 2 and 4). The plots in columns 1 and 2 were obtained from the responses to the first click in the click train (Click \#1), while the plots in columns 3 and 4 were obtained from the responses to the 10\textsuperscript{th} click in the click train.\vspace{-22pt}}
	\label{fig:6}
\end{figure} 

\textbf{Evaluation stimuli}: A click train with a repetition rate of 20~Hz and duration of 0.5~s (10 clicks) was generated \cite{Dau2000}. The clicks had a positive polarity and each click had a duration of 80~$\mu$s. The first click was set to start after 10~$\mu$s. The level of the click was varied between 60 and 100~dB~peSPL. \vspace{4pt}

\textbf{Hearing profiles}: Three hearing profiles were considered: NH (Flat00, 13-3-3), HI (Slope35, 13-3-3), and HSR (Flat00, 13-0-0). (1) The \textbf{NH} profile uses a normal-hearing cochlear model with no synaptopathy. (2) The \textbf{HI} profile simulates a high-frequency sloping audiogram with 0 dB HL at 1~kHz and 35 dB HL at 8~kHz and no synaptopathy. (3) the \textbf{HSR} profile uses a normal hearing cochlear model and simulates a loss of medium- and low-spontaneous rate neurones. A graphical representation of the cochlear-gain-loss profiles Flat00 and Slope35 can be found in \cite{Verhulst2016}.\vspace{4pt}

\textbf{Results}: The ABRs presented in this section were used to reproduce the simulation results shown in Fig.~6 of the original model paper, where ABR latencies and amplitudes derived from a single click were reported (``Click \#1''). For ease of comparison with the current results, the panels in Fig.~\ref{fig:6} are labelled as A, B, and C (here from top to bottom), respectively. In Fig.~\ref{fig:6}, the left-most panels (column 1) reproduce the ABR results from the original model paper (using \textbf{model v1.1}), while the middle panels (column 2) show the ABRs to Click \#1 but then using the \textbf{model v1.2} implementation. However, given that the original aim of the model was to mimic experimental findings to reference ABRs recorded to click trains, the model paper should have rather used click trains in its evaluation/calibration than using Click \#1. The experimental data labelled in the model paper as `D03' (reported in \cite{Dau2003} but originally collected in \cite{Dau2000}) used a repetition rate of 20 Hz. To provide better experimental consistency, new simulation results were generated with ABRs derived from the 10\textsuperscript{th} click in the train using model v1.1 and v1.2 in panels 3 and 4, respectively. This new approach ensures that all AN time constants have reached steady-state as can be expected experimentally.
The ABRs in the right-most panels (column 4) should hence be more directly comparable to the experimental results in \cite{Dau2003} and other experimental references reported in the model paper.\vspace{4pt}

\textbf{Discussion: Simulations with Click \#1}: Shorter latencies and larger W-I and W-V amplitudes were obtained for \textbf{model v1.2} compared to \textbf{model v1.1}. This was an expected consequence of the adopted calibration procedure, which was based on later clicks rather than to the first click of the click train, which was not affected by AN adaptation. Due to the larger W-I and W-V amplitudes in \textbf{v1.1}, the derived latencies were shorter in \textbf{v1.2} than originally, and this effect was more pronounced in the HI profile  (purple line) and almost negligible in the NH (green lines) and HSR profiles (red dashed lines).\vspace{4pt}

\textbf{Discussion: Simulations with Click \#10}: For the same model version, i.e., panel~3 compared to panel~1 (\textbf{model v1.1}), and panel 4 with panel 2 (\textbf{model v1.2}), the obtained ABR amplitudes (see Fig.~\ref{fig:6}C) using Click 10\textsuperscript{th} were lower, which was an expected result. The latencies were also shorter, especially for the lower level clicks (60 and 70 dB peSPL).\vspace{20pt}

\textbf{Discussion: Simulations with Click \#1 (model v1.1, panel 1) and Click \#10 (model 1.2, panel 4)}: Both latencies and amplitudes had a similar range. The results with Click \#10 (panel 4) reflect more closely the experimental data \cite{Dau2000} which were collected to click trains with a repetition rate of 20~Hz. Note that the peak amplitudes at 100 dB peSPL in panel 1 coincide with the normative data (0.61$\mu$V for W-V; 0.30$\mu$V for W-I), whereas the amplitudes are slightly higher in panel~4. This is a consequence of the lower repetition rate of the clicks (20~Hz) and, hence, the longer neuronal firing recovery compared to a click presentation at 11.1~Hz, which was used in the model calibration.

\newpage
\vspace{-8pt}
\subsection{AM tones: Simulating EFRs}

\begin{figure} [!b]
	\centering
	\subfigure[\textbf{Simulated EFR amplitudes} obtained using Eq. (\ref{eqn:EFR}) for a 4-kHz AM tone with f\textsubscript{mod}=98~Hz (85\% modulation depth) presented to the model at different levels. Three hearing profiles are shown: NH (green squares), HI (purple circles), and HSR (red dashed diamonds). These panels are similar to Fig. 7A from \cite{Verhulst2018a}.]
	{
	\includegraphics[width=0.39\textwidth]{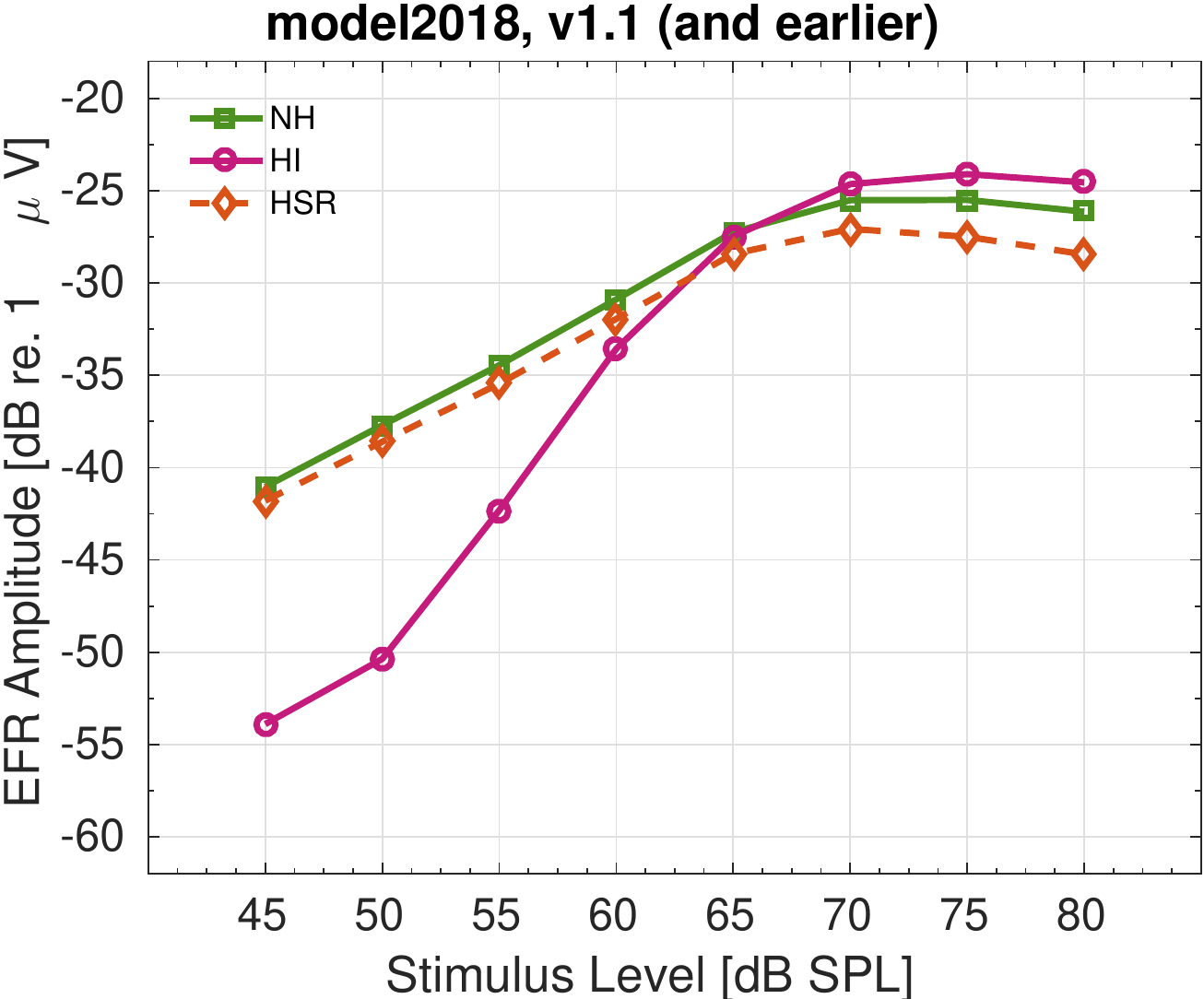}
	\includegraphics[width=0.39\textwidth]{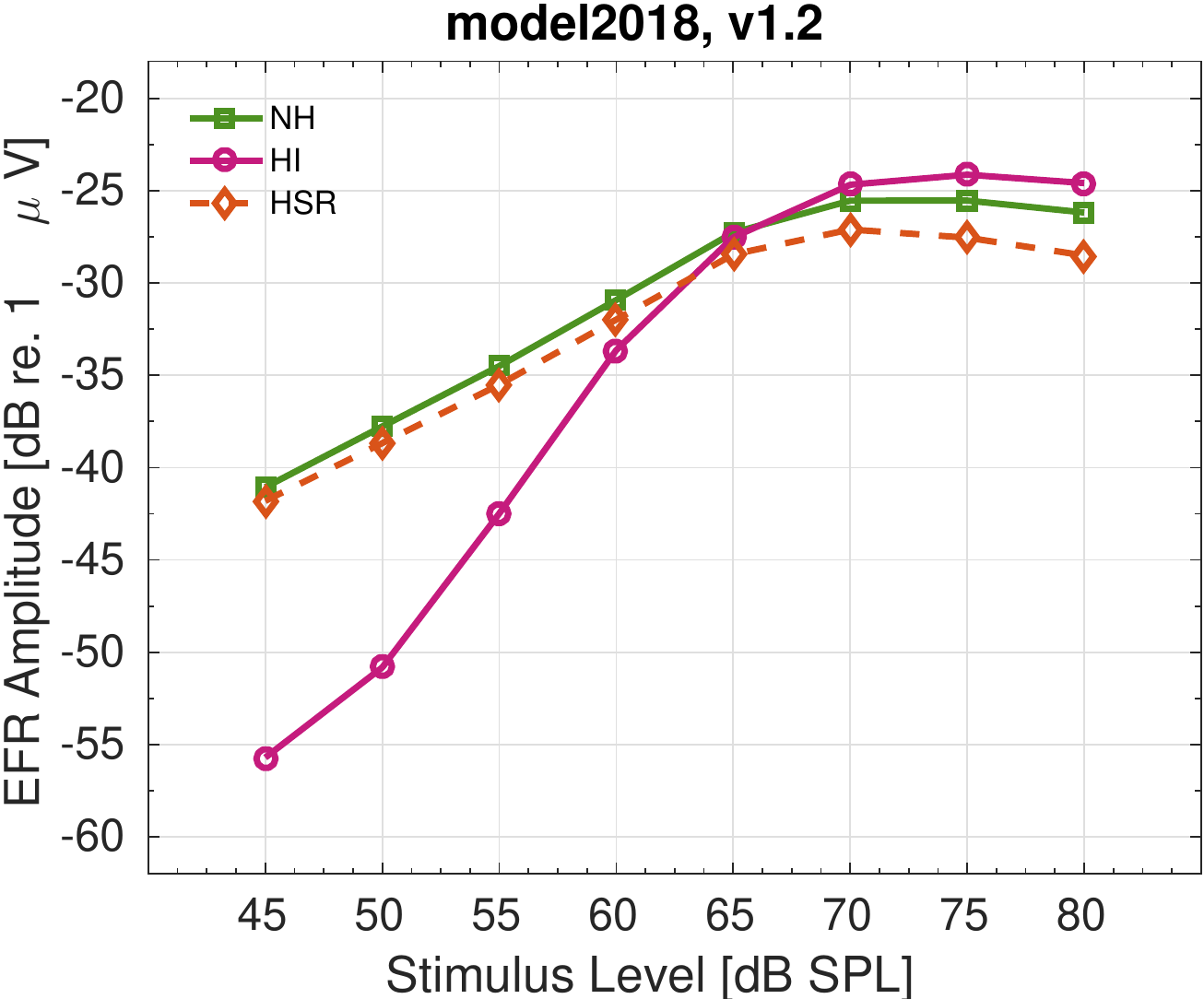}
	}
	\subfigure[\textbf{Simulated EFR amplitudes} for the same 4-kHz tones of panel (a) obtained with Eq. (\ref{eqn:EFR}) but using $r\textsubscript{EFR}$ (Eq. \ref{eqn:rEFR}) with only on-frequency (filled markers) or off-frequency contributions (open markers) for the NH (green squares) and HSR profiles (red dashed diamonds). The on-frequency and off-frequency contributions were obtained from the CF channels around 4~kHz (channels 100 to 123) and 8~kHz (channels 30 to 54), respectively. See the text for further details. These panels are similar to Fig. 7B from~\cite{Verhulst2018a}.]
	{
	\includegraphics[width=0.39\textwidth,trim=0 0 0 0.6cm,clip=true]{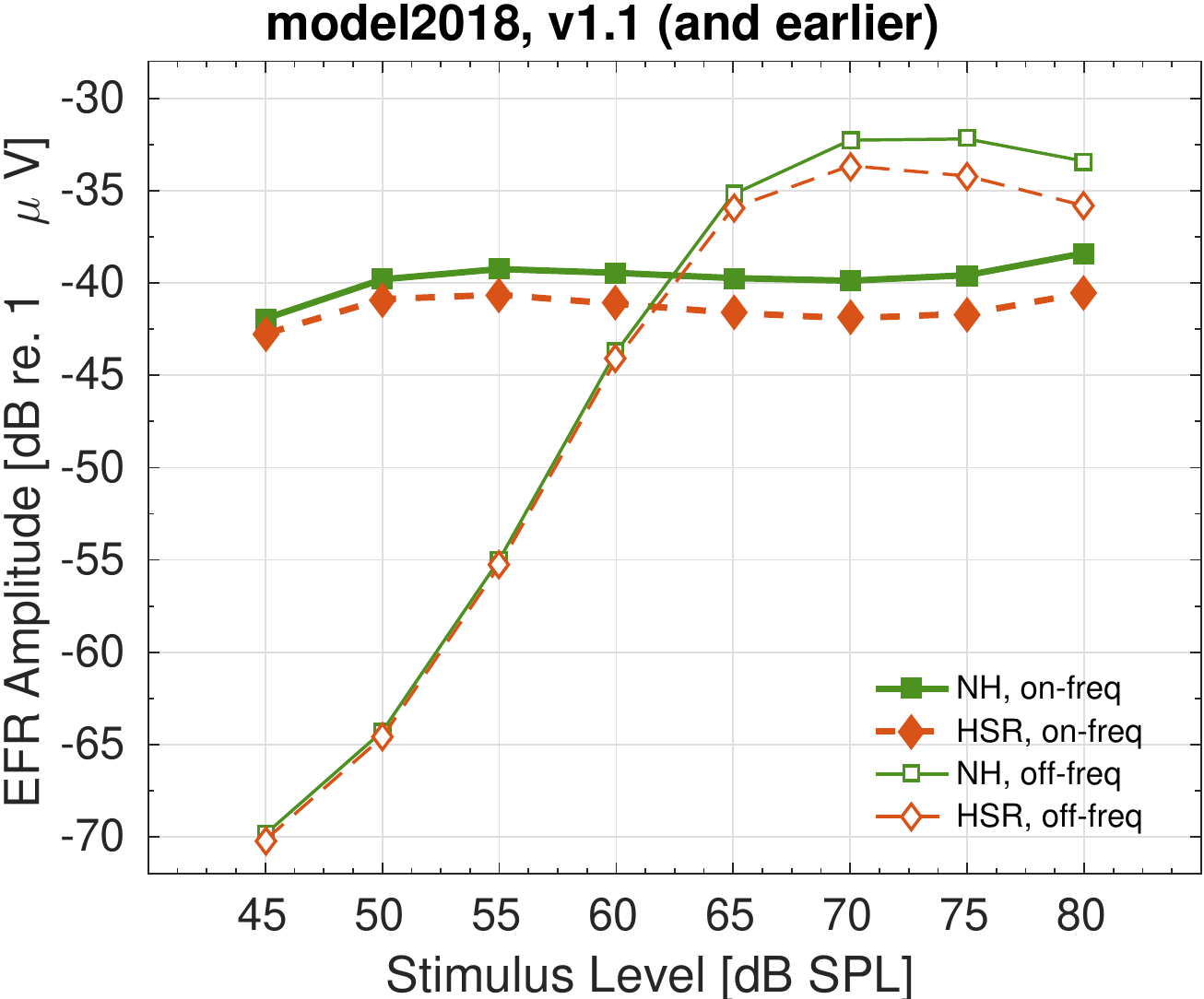}
	\includegraphics[width=0.39\textwidth,trim=0 0 0 0.6cm,clip=true]{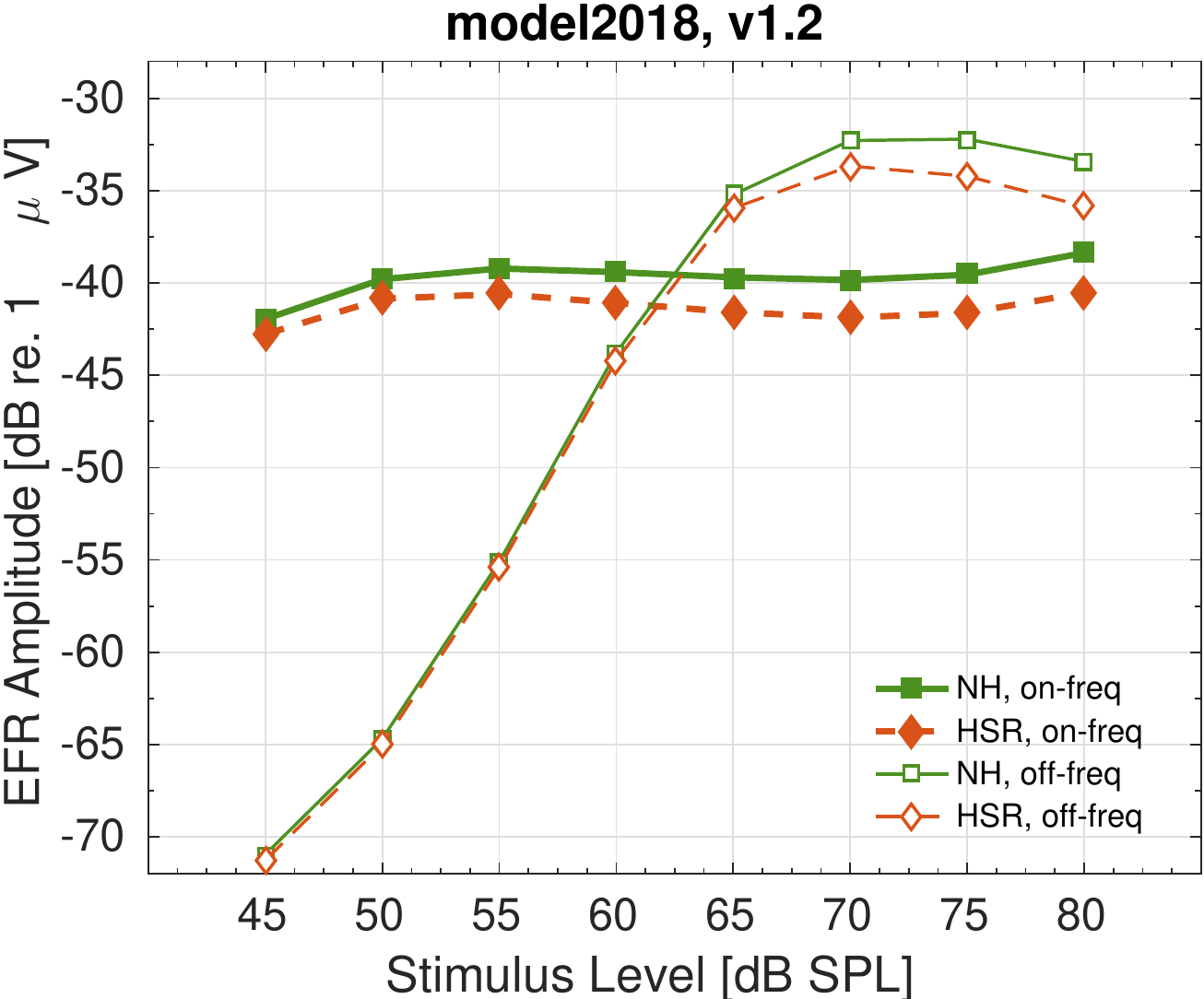}
	}
	\subfigure[Simulated $r\textsubscript{EFR}$ (Eq.~\ref{eqn:rEFR}) in time domain using the 70-dB tone.]
	{
	\includegraphics[width=0.39\textwidth,trim=0 0 0 0.6cm,clip=true]{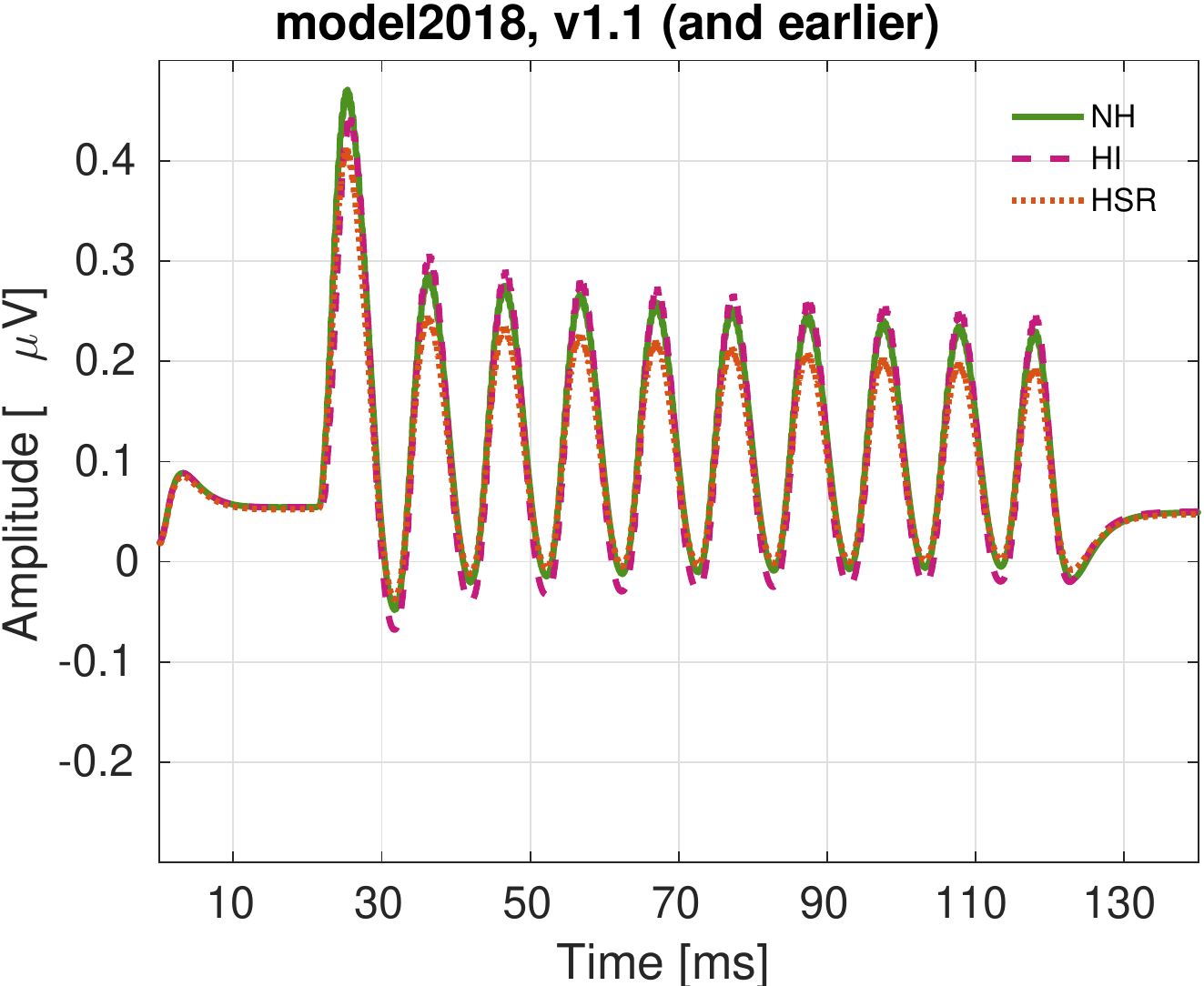}
	\includegraphics[width=0.39\textwidth,trim=0 0 0 0.6cm,clip=true]{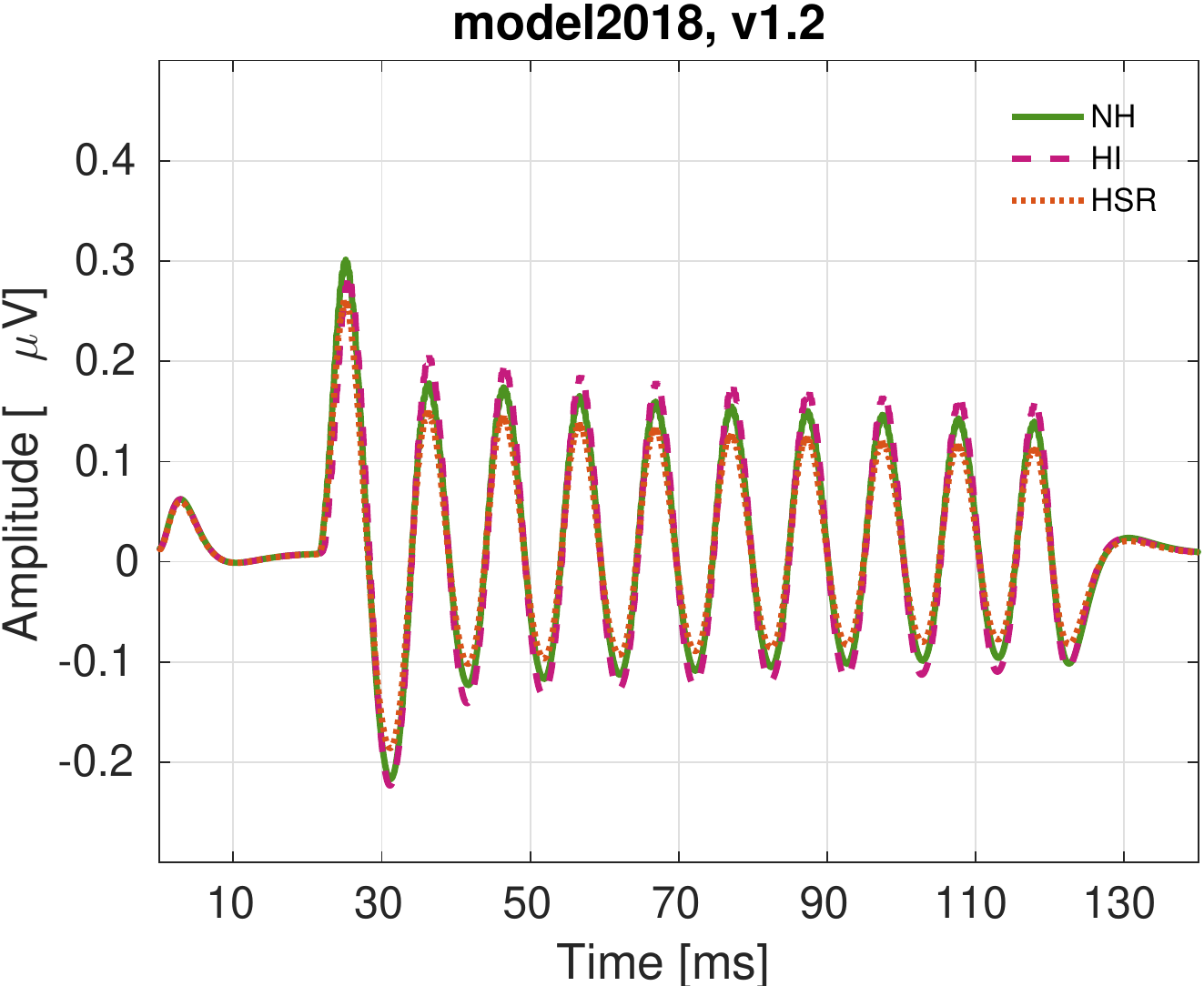}
	}
	\vspace{-12pt}
	\caption{Simulations using \textbf{model v1.1} (left panels) and \textbf{model v1.2} (right panels).\vspace{-20pt}}
	\label{fig:7}
\end{figure}

\textbf{Stimuli}: A 100-ms pure tone with a carrier frequency of 4~kHz was modulated at a rate of 98~Hz using 85\% modulation depth. An up-down cosine ramp of 2.5~ms was applied. The sound was preceded by 20~ms of silence before the stimulus was fed into the model.\vspace{4pt}

\textbf{EFR assessment}: Envelope-following responses (EFRs) were obtained by adding the simulated AN responses, CN, and IC responses as follows:

\vspace{-8pt}
\begin{equation}
	r\textsubscript{EFR}(t)=A\textsubscript{W-I}\cdot r\textsubscript{AN}(t)+A\textsubscript{W-III}\cdot r\textsubscript{CN}(t)+A\textsubscript{W-V}\cdot r\textsubscript{IC}(t) 
	\label{eqn:rEFR}
\end{equation}
Broadband population responses (Fig.~\ref{fig:7}, top panels) were obtained by adding the CF contributions from all simulated frequency bins (401 channels), between 112 Hz (channel 401) and 12~kHz (channel 1). On-frequency and off-frequency EFRs (Fig.~\ref{fig:7}, middle panels) were assessed by summing up the CF contributions within one-third octave-band around 4~kHz (channel 112, channels 100 to 123) and 8~kHz (channel 42, channels 30 to 54), respectively. The simulated population responses (4000-samples long: 200~ms at f$_\textsubscript{s,ABR}=20$~kHz) were converted to the frequency domain by applying a 4000-point FFT (Fig.~\ref{fig:7}, top and middle panels) and were expressed in dB re. 1 $\mu$V:
\begin{equation}
	\mbox{EFR}\textsubscript{magnitude}=20\cdot \log_{10}{\left( \frac{\max{\left(\|\mbox{FFT}(r\textsubscript{EFR})\|\right)/N}}{1\cdot 10^{-6}}\right)} \mbox{\hspace{20pt}[dB re.~}1~\mu\mbox{V]}
	\label{eqn:EFR}
\end{equation}


\textbf{Results}: The corresponding model responses are shown in Fig.~\ref{fig:7}. \vspace{4pt} 

\textbf{Discussion}: The results shown in panels (a) and (b) of Fig.~\ref{fig:7} are similar for both model versions. However, the time domain representation of the 70-dB envelope-following response of panel (c) clearly shows more inhibition (negative voltages) for \textbf{model v1.2}. This is a consequence of the corrected inhibition/excitation balance of the IC model stage.

\vspace{-10pt}
\printbibliography

\end{document}